\documentclass[a4paper,11pt]{article}
\usepackage{latexsym}
\usepackage{amssymb}
\usepackage{amsfonts}
\usepackage{amsmath}
\newcommand{\bol}[1]{\boldsymbol {#1}}
\newcommand{\sss}[1]{\scriptscriptstyle {#1}}
\newcommand{\va}{\varphi}

\begin{document}
\author{Nicola Grillo\thanks{\texttt{grillo@physik.unizh.ch}} \\  \\
{\textit{Institut f\"ur Theoretische Physik, Universit\"at Z\"urich}} \\
{\textit{Winterthurerstrasse 190, CH-8057 Z\"urich, Switzerland} }}
\title{Quantization of the Graviton Field, Characterization of 
the Physical Subspace and Unitarity in Causal Quantum Gravity}
\maketitle
\begin{abstract}
\noindent
In this paper we lay the foundations of causal quantum gravity,
\emph{i.e.} of a quantum theory of self-interacting symmetric  massless rank-2 tensor 
gauge fields, the {\textsl{gravitons}}, on flat space-time, in the framework of causal
perturbation theory. 
The causal inductive construction of the $S$-matrix for quantum gravity
leads to a very satisfactory treatment of the ultraviolet problem.
Here we concentrate on some main fundamental issues 
that concern the quantization of the gravitational interactions:
the quantization of the free graviton field, the r\^ole of the fermionic
ghost vector fields in preserving perturbative gauge invariance, 
the construction of the Fock space for the graviton and the consequent
characterization of the subspace for the physical graviton states.
This last point is necessary, together with perturbative gauge invariance,
in order to prove unitarity of the $S$-matrix restricted to the 
physical subspace.
\vskip 1cm
{\bf PACS numbers:} 0460, 1110
\vskip 7mm
{\bf Keywords:} Quantum Gravity
\vskip 7mm
{\bf Preprint:} ZU-TH 35/1999
\end{abstract}
\newpage
\tableofcontents
\newpage
\section{Introduction}
\setcounter{equation}{0}

The central aspect of this work is the understanding of some fundamental problems that arise in the quantization process
of a rank-2 tensor  field and in the perturbative construction of the scattering matrix for the corresponding  quantum
field theory: quantum gravity (QG), by which we mean a quantum field theory of self-interacting massless rank-2 symmetric
tensor gauge  fields on flat Minkowski background metric.

A review of such a broad subject, with all the implications that it involves, cannot be the theme of a short introduction,
we therefore refer to much more complete works:~\cite{dew5},~\cite{isha2}.

One of the possible derivation of Einstein's theory of gravitation is the field-theoretic, non-geometrical route
which essentially identifies massless spin-2 fields with carriers of the gravitational interaction.
Numerous papers discuss this 
approach:~\cite{gu2},~\cite{fey1},~\cite{ogipu},~\cite{wyss} and~\cite{mes2} and references therein.

For the implementation of this approach on a quantum level, namely the implementation of quantum gravity
as a Poincar\'e covariant local quantum field theory with a considerable gauge arbitrariness, two methods
will be used: the Epstein--Glaser inductive construction of the $S$-matrix perturbation series~\cite{eg}
and the concept of `perturbative quantum operator gauge invariance', borrowed from the case of
non-Abelian gauge theories~\cite{nonab}.

The first method provides an elegant and efficacious way of dealing with the ultraviolet problem of quantum gravity
taking causality as  cornerstone in the construction of the $S$-matrix. Although in this
paper we will not concentrate on this  aspect  (loop calculations will be the subject of a forthcoming work),
a short  introduction to the causal Epstein--Glaser method is given.

The second method formulates perturbative gauge invariance by means of a gauge charge $Q$ which 
requires the introduction of the ghost-graviton  coupling and allows us
to construct explicitely the Fock space of  physical graviton  states.
This is a very important issue because a rank-2 tensor field has more independent degrees of freedom than
a massless spin-2 particle.
In addition, we are  able to prove perturbative  unitarity of the $S$-matrix on the physical subspace.

A different formulation of the graviton quantization can be found in~\cite{grigo}, whereas 
results obtained in causal quantum gravity by the author are reviewed in~\cite{gri2}.

The paper is organized as follow: in the next section we briefly introduce 
the causal method in $S$-matrix perturbation theory, in Sec.~\ref{sec:gr} we
derive from classical general relativity the inputs that are needed for the causal construction
of the $S$-matrix for QG, only in this section we resort to classical general relativity.
In Sec.~\ref{sec:gauge} we introduce the concept of perturbative
quantum operator gauge invariance.
In Sec.~{\ref{sec:quant} we discuss classical and quantized tensor fields and give an
explicit representation for the latter which enables us to construct the graviton Fock space
$\mathcal{F}$. The explicit construction of the physical subspace $\mathcal{F}_{phys}\subset\mathcal{F}$ 
is carried out in Sec.~\ref{sec:physsub} after quantization of the ghost and anti-ghost fields.
The analysis of the Krein conjugation on $\mathcal{F}$, in Sec.~\ref{sec:krein}, entails pseudo-unitarity
instead of unitarity. But unitarity can be restored, in Sec.~\ref{sec:poinc}, between  graviton states
under Poincar\'e transformations  and
in Sec.~\ref{sec:unit} for the  $S$-matrix, although for its restriction on the physical subspace.

We use the unit convention: $\hbar=c=1$, Greek indices $\alpha,\beta,\ldots$
 run from $0$ to $3$, whereas Latin indices $i,j,\ldots$ run from $1$ to $3$.

\section{\boldmath$S$-Matrix Causal Perturbation Theory}\label{sec:smatr}

\setcounter{equation}{0}

Since in this paper the causal  method~\cite{eg} is not used  in its full strength, but it serves only
as a motivation and starting point for further investigations, we give only a concise review 
of the causal approach to QFT, for a detailed exposition, see \cite{scha}, \cite{aste}.

We consider the $S$-matrix, being a formal power series in the coupling constant, as a sum of 
smeared operator-valued distributions of the following form:
\begin{equation}
S(g)={\mathbf 1}+\sum_{n=1}^{\infty}\frac{1}{n!} \int\!\! d^{4}x_{1}
\ldots d^{4}x_{n}\, T_{n}(x_{1},\ldots, x_{n})\,g(x_{1})\cdot\ldots\cdot g(x_{n})\,  ,
\label{eq:1.1}
\end{equation}
where $g$ is a Schwartz test function ($g\in\mathcal{S}(\mathbb{R}^{4})$) which switches adiabatically the
interaction and provides a natural infrared cutoff in the long-range part 
of the interaction.

The $S$-matrix maps the asymptotically incoming free fields on the outgoing ones and it is possible
to express it by means of free fields. Interacting quantum fields are not used in the causal approach.

The $n$-point operator-valued distribution $T_{n}$ is a  
well-defined  `renormalized' time-ordered product expressed in terms of Wick monomials of free fields. 
$T_{n}$ is constructed inductively from the first order $T_{1}(x)$, which corresponds to the usual interaction 
Lagrangian in terms of free fields, and from the lower orders $T_{j}$, 
$j=2,\ldots,n-1$ by means of Poincar\'e covariance and causality. 

Causality, if correctly incorporated, leads directly to the `renormalized' perturbation series for 
the $S$-matrix which is ultraviolet (UV) finite and cutoff-free in every order. 

The construction of $T_{n}$ requires some care: if it were simply given by the usual time-ordering
\begin{equation}
\begin{split}
T_{n}(x_{1},\ldots,x_{n})&=T\{T_{1}(x_{1})\ldots T_{1}(x_{n})\}\\ 
                         &=\sum_{\pi \in \sigma_{n}}
                    \Theta (x^{0}_{\pi(1)}-x^{0}_{\pi(2)})\ldots\Theta(x^{0}_{\pi(n-1)}-x^{0}_{\pi(n)})\cdot \\    
                         &\qquad \qquad\qquad \cdot T_{1}(x_{\pi(1)})\ldots T_{1}(x_{\pi(n)})\, ,
\end{split}
\label{eq:1.2}
\end{equation}
then UV-divergences would appear. 

For the causal inductive  construction, the equations of motion of the free 
quantum fields are needed together with their commutation rules expressed by  the causal Jordan-Pauli 
distribution $D_{m}(x)=D_{m}^{\sss (+)}(x)+D_{m}^{\sss (-)}(x)$ for a field with mass $m$.

Then, an auxiliary distribution $D_{n}(x_{1},\ldots ,x_{n})$ is constructed with the help of 
$T_{1},\ldots , T_{n-1}$ by carrying out all the possible contractions between field operators appearing
in $T_{1},\ldots , T_{n-1}$ using Wick's lemma, so that $D_{n}$ has the following form
\begin{equation}
\begin{split}
D_{n}(x_{1},\ldots ,x_{n})&=R_{n}'(x_{1},\ldots ,x_{n})-A_{n}'(x_{1},\ldots ,x_{n})\\
                      &=\sum_{k} :\!\mathcal{O}_{k}(x_{1},\ldots ,x_{n})\!:\, 
                     d_{n}^{\sss {\left[ k \right]}}(x_{1}-x_{n},\ldots ,x_{n-1}-x_{n})\,  ,
\label{eq:1.3}
\end{split}
\end{equation}
where $:\!{\mathcal O}_{k}(x_{1},\ldots ,x_{n})\!:$ represents a normally ordered product of free
field operators and $d_{n}^{\sss {\left[ k \right]}}(x_{1}-x_{n},\ldots , x_{n-1}-x_{n})$ is a 
C-number distribution that contains well-defined products of positive and negative frequency 
parts of $D_{m}(x_{j}-x_{n})$ with $j=1,\ldots ,n-1$.
Because of translation invariance, $d_{n}^{\sss {\left[ k \right]}}$ depends only on relative 
coordinates with respect to $x_{n}$; Eq.~(\ref{eq:1.3}) contains tree, loop and vacuum contributions.

The most important property of $D_{n}$ is causality:
\begin{equation}
\hbox{supp}\big(D_{n}(x_{1},\ldots ,x_{n})\big)\subseteq \Gamma^{+}_{n-1}(x_{n}) \cup \Gamma^{-}_{n-1}(x_{n})\,  ,
\label{eq:1.4}
\end{equation}
where
\begin{equation}
\Gamma^{\pm}_{n-1}(x_{n})=\{ (x_{1},\ldots ,x_{n})\in \mathbb{R}^{4n}\mid\, x_{j}\in (x_{n}+\overline{V^{\pm}}),\,
\forall j=1,\ldots ,n-1\}\, .
\label{eq:1.5}
\end{equation}
Only the numerical distributions are responsible for the support properties. 
In order to obtain 
$T_{n}(x_{1},\ldots ,x_{n})$ we have to split correctly the $D_{n}$-distribution into a retarded part, 
$R_{n}$, and an advanced part, $A_{n}$, with
\begin{equation*}
\hbox{supp}\big(R_{n}(x_{1},\ldots ,x_{n})\big)\subseteq \Gamma^{+}_{n-1}(x_{n})\quad\mathrm{and}\quad 
\hbox{supp}\big(A_{n}(x_{1},\ldots ,x_{n})\big)\subseteq \Gamma^{-}_{n-1}(x_{n})\, .
%\label{eq:1.6}
\end{equation*}
This  operation affects only the numerical distributions $d_{n}^{\sss {\left[ k \right]}}$ and obviously 
the critical point for the splitting lies in the coincidence point
\begin{equation}
\Gamma^{+}_{n-1}(x_{n}) \cap \Gamma^{-}_{n-1}(x_{n})=\Delta_{n}=\{x_{1}=x_{2}=\ldots =x_{n}\}\, .
\label{eq:1.7}
\end{equation}
The correct treatment of this point constitutes the key  to control the UV-behaviour of the $n$-point 
distribution. Indeed, the origin of the UV-divergences lies in products of Feynman propagators with 
coincident arguments  in loop graphs, because time-ordering cannot be done simply by multiplying 
(singular) distributions by discontinuous $\Theta$-distributions, as in Eq.~(\ref{eq:1.2}), a procedure which is usually 
ill-defined, but if it is carefully done by first multiplying with a $C^{\infty}$-function and then 
performing the limit to step function, then the expression for $T_{n}$ remains well-defined and finite.

The measure of the behaviour of the numerical distribution $d_{n}^{\sss {\left[ k \right]}}$ near this point is 
encoded in the number $\omega(d_{n}^{\sss {\left[ k \right]}})$   called `singular order'.

The  splitting of the numerical distribution $d_{n}^{\sss {\left[ k \right]}}$ must therefore be  
accomplished according to the correct singular order $\omega(d_{2}^{\sss {\left[ k \right]}})$, otherwise
UV-divergences would appear. If $\omega < 0$, then the splitting is trivial and agrees with the standard 
time-ordering and 
we recover the Feynman rules. On the other side, if $\omega\ge 0$, then the splitting is non-trivial and 
non-unique. For a general  retarded part we obtain:
\begin{multline}
d_{n}^{\sss{\left[ k \right]}}(x_{1}-x_{n},\ldots ,x_{n-1}-x_{n})\longrightarrow   r_{n}^{\sss {\left[ k \right]}}
(x_{1}-x_{n},\ldots ,x_{n-1}-x_{n}) +\\
  +\sum_{|a| =0}^{\omega(d_{n})} C_{a,\sss{\left[k\right]} }\, D^{a}\, \delta^{\sss (4(n-1))}
(x_{1}-x_{n},\ldots ,x_{n-1}-x_{n})\,  ,
\label{eq:1.8}
\end{multline}
where a special retarded part $r_{n}^{\sss {\left[ k \right]}}$, the so-called central splitting solution,
is obtained in momentum space from $\hat{d}_{n}^{\sss {\left[ k \right]}}$ by means of a  dispersion integral:
\begin{equation}
\hat{r}_{n}^{\sss{\left[ k \right]}}(p)=\frac{i}{2\pi}\int_{-\infty}^{\infty}\!\! dt \frac{\hat{d}_{n}^{\sss
 {\left[ k \right]}}(tp)}{\left( t-i0\right)^{\omega +1}\left( 1-t+i0\right)}\,,
\quad  p=(p_{1},\ldots ,p_{n-1}),\ p_{j}\in \overline{V^{+}} \, .
\label{eq:1.9}
\end{equation}
Eq.~(\ref{eq:1.8}) contains a local ambiguity in the normalization: the $C_{a,\sss{\left[k\right]}}$'s are undetermined 
finite normalization constants, which multiply terms with local support. This freedom in the normalization has to 
be restricted by further physical conditions, \emph{e.g.} Lorentz covariance, pseudo-unitarity
(see Sec.~\ref{sec:unit}), existence of the adiabatic limit $g\to 1$ and gauge invariance (see Sec.~\ref{sec:ghost}). 

Finally, $T_{n}$ is given by
\begin{multline}
T_{n}(x_{1},\ldots ,x_{n})+N_{n}(x_{1},\ldots ,x_{n})=R_{n}(x_{1},\ldots ,x_{n})-R_{n}'(x_{1},\ldots ,x_{n})\\
                  =\sum_{k} :\!{\mathcal O}_{k}(x_{1},\ldots ,x_{n})\!: \,  
                   t_{n}^{\sss {\left[ k \right]}} (x_{1}-x_{n},\ldots , x_{n-1}-x_{n})+\\
            +\sum_{k} :\!{\mathcal O}_{k}(x_{1},\ldots ,x_{n})\!: \, \Big\{ \sum_{|a| =0}^{\omega} C_{a,\left[k\right]}\, 
                    D^{a}\, \delta^{\sss (4(n-1))}(x_{1}-x_{n},\ldots ,x_{n-1}-x_{n})\Big\} \, ,
\label{eq:1.10}
\end{multline}
in which we artificially separate local from non-local contributions for later use.

Most of the $S$-matrix properties are indeed translated into conditions that the $T_{n}$'s and the $N_{n}$'s  must
satisfy, for example gauge invariance and unitarity.

The advantages of the causal scheme are that it leads directly to a `renormalized' perturbative
expansion for the $S$-matrix without using a cutoff. It makes possible, to a given order in the coupling 
constant, to compute UV finite amplitudes for all  processes and it does not rely on the Lagrangian 
approach with interacting quantum fields.

\section{From General Relativity to Quantum Gravity}\label{sec:gr}

\setcounter{equation}{0}

The causal inductive construction of the n-point operator valued distributions $T_{n}$ that appear in the expansion of 
the $S$-matrix can be accomplished provided some basic assumptions about the free fields and their couplings are made.
For QG in the causal framework, we need the equation of motion of the free graviton field in a fixed gauge, the 
commutation rule between two graviton fields at different space-time points and the first order graviton coupling.

\subsection{Classical General Relativity in Lagrangian Form}

Since we are interested in a quantum theory of Einstein's general relativity, we start by considering the 
Hilbert--Einstein Lagrangian density (without cosmological constant)
\begin{equation}
\mathcal{L}_{\sss HE}={\frac{- 2}{\kappa ^ 2}}\sqrt{-g} R\, ,
\label{eq:2.1}
\end{equation}
with the notation:
\begin{equation}
\begin{split}
R_{\mu\nu}&=R^{\sigma}_{\: \mu\sigma\nu}= \Gamma_{\mu\nu\ ,\alpha}^{\alpha}-\Gamma_{\mu\alpha\ ,\nu}^{\alpha}+ 
                                      \Gamma_{\mu\nu}^{\beta} \Gamma_{\beta\alpha}^{\alpha}-
                                      \Gamma_{\mu\alpha}^{\beta}\Gamma_{\beta\nu}^{\alpha}\, ,\\
\Gamma_{\mu\nu}^{\alpha}&=\frac{1}{2}g^{\alpha\beta}\left( g_{\beta\mu , \nu}+g_{\beta\nu , \mu}-g_{\mu\nu , \beta}\right)\, ,\qquad 
R=g^{\mu\nu}R_{\mu\nu}\,,
\label{eq:2.2}
\end{split}
\end{equation}
and $\kappa^2=32\pi G$, where $G$ is the Newton's constant. The variation of the source-free Hilbert--Einstein action 
\begin{equation}
\mathcal{A}_{\sss HE}={\frac{- 2}{\kappa ^ 2}}\int \!\! d^4x\, \sqrt{-g}\, R\, ,
\label{eq:wirkung}
\end{equation}
with respect to the metric tensor $g_{\mu\nu}$ leads to the vacuum Einstein field equations
\begin{equation}
G^{\mu\nu}:=R^{\mu\nu}-\frac{1}{2}g^{\mu\nu}R=0\, .
\label{eq:2.4}
\end{equation}
We rewrite $\mathcal{L}_{\sss HE}$ in term of the Goldberg variable
\begin{equation}
\tilde{g}^{\mu\nu}:=\sqrt{-g}\: g^{\mu\nu}\, ,
\label{eq:2.5}
\end{equation}
and, disregarding divergence terms that can be integrated away, we obtain
\begin{equation}
\mathcal{L}_{\sss HE}=\frac{1}{2\kappa^2}\left[+\tilde{g}^{\rho\sigma}\tilde{g}_{\lambda\alpha}\tilde{g}_{\kappa\tau}
                                   -\frac{1}{2}\tilde{g}^{\rho\sigma}\tilde{g}_{\alpha\kappa}\tilde{g}_{\lambda\tau}
                                   -2\tilde{g}_{\alpha\tau}\eta^{\sigma}_{\kappa}\eta^{\rho}_{\lambda} \right]
                             \tilde{g}^{\alpha\kappa}_{\: ,\rho}\tilde{g}^{\lambda\tau}_{\: ,\sigma}\, ,
\label{eq:2.6}
\end{equation}
where $\eta^{\mu\nu}=\mathrm{diag}(1,-1,-1,-1)$ is the flat space-time metric tensor. 

After having defined the symmetric tensor field $h^{\mu\nu}(x)$, the `graviton' field, through the expansion 
of the Goldberg variable
\begin{equation}
\tilde{g}^{\mu\nu}(x)=\eta^{\mu\nu}+\kappa\, h^{\mu\nu}(x)\, ,
\label{eq:gtilde}
\end{equation}
in the flat geometry, we insert~(\ref{eq:gtilde}) into~(\ref{eq:2.6}) and obtain a non-terminating
expansion of the Hilbert--Einstein Lagrangian density
\begin{equation}
\mathcal{L}_{\sss HE}=\sum_{j=0}^{\infty}\, \kappa^{j}\, \mathcal{L}_{\sss HE}^{\sss{(j)}}\, ,
\label{eq:2.7}
\end{equation}
where $\mathcal{L}_{\sss HE}^{\sss{(j)}}$ represents an interaction involving $j+2$ gravitons.

\subsection{Linearized Free Theory}

For convenience of notation, the trace of the graviton field is written as $h=h^{\gamma}_{\  \gamma}$ and 
all Lorentz indices are written as superscripts whereas the derivatives are written as subscripts. 
All indices occurring twice are contracted by the Minkowski metric $\eta^{\mu\nu}=\mathrm{diag}(1,-1,-1,-1)$.

The lowest order $\mathcal{L}_{\sss HE}^{\sss{(0)}}$ is quadratic in the field $h^{\mu\nu}(x)$:
\begin{equation}
\mathcal{L}_{\sss HE}^{\sss{(0)}}=\frac{1}{2}h^{\mu\nu}_{\: ,\rho}h^{\mu\nu}_{\: ,\rho}-\frac{1}{4}h_{,\rho}h_{,\rho}
            -h^{\mu\nu}_{\: ,\rho}h^{\mu\rho}_{\: ,\nu}\, ,
\label{eq:2.8}
\end{equation}
from which, by means of the Euler--Lagrange variations with respect to $h^{\mu\nu}$
\begin{equation}
\frac{\partial \mathcal{L}_{\sss HE}^{\sss{(0)}}}{\partial h^{\mu\nu}}-\partial_{\rho}
    \frac{\partial \mathcal{L}_{\sss HE}^{\sss{(0)}}}{\partial h^{\mu\nu}_{\: ,\rho}}=0\, ,
\label{eq:2.9}
\end{equation}
we obtain the free equation of motion for $h^{\mu\nu}$:
\begin{equation}
\Box h^{\mu\nu}-\frac{1}{2}\eta^{\mu\nu}\Box h - h^{\mu\rho}_{,\rho\nu}- h^{\nu\rho}_{,\rho\mu}=
\frac{2}{\kappa}\, R^{\mu\nu}_{\sss{L}}=0 \, ,
\label{eq:2.10}
\end{equation}
where $R^{\mu\nu}_{\sss{L}}$ is the linearized, \emph{i.e.} up to order $\kappa$, Ricci tensor.

The equation of motion~(\ref{eq:2.10}) and the free Lagrangian~(\ref{eq:2.8}) are invariant under the gauge transformation
\begin{equation}
h^{\mu\nu}\longrightarrow {h'}^{\mu\nu}=h^{\mu\nu}+f^{\mu}_{,\nu}+f^{\nu}_{,\mu}-\eta^{\mu\nu}f^{\gamma}_{\  ,\gamma}\, ,
\label{eq:2.11}
\end{equation}
if the vector field $f^{\mu}$ satisfies the wave equation $\Box f^{\mu}=0$.

The classical gauge transformation~(\ref{eq:2.11}) corresponds to the linearized general covariance of $g_{\mu\nu}(x)$:
under a general coordinate transformation of the form ${x'}^{\mu}={x'}^{\mu}(x)$,
the metric tensor transforms according to
\begin{equation}
g_{\mu\nu}(x)=\frac{\partial {x'}^{\sigma}}{\partial x^{\mu}}\, \frac{\partial {x'}^{\rho}}{\partial x^{\nu}}
    g'_{\rho\sigma}(x')\, .
\label{eq:2.12}
\end{equation}
Expanding $g_{\mu\nu}(x)=\eta_{\mu\nu}+\kappa\, \phi_{\mu\nu}(x)$ and considering a coordinate transformation of the form
${x'}^{\mu}=x^{\mu}+f^{\mu}$, we find that  terms in Eq.~(\ref{eq:2.12}) that are linear in the fields $\phi^{\mu\nu}$ 
and $f^{\mu}$ are related by the transformation
\begin{equation}
\phi^{\mu\nu}\longrightarrow {\phi'}^{\mu\nu}=\phi^{\mu\nu}+f^{\mu}_{,\nu}+f^{\nu}_{,\mu}\, .
\label{eq:2.13}
\end{equation}
Since the tensor fields $\phi^{\mu\nu}$ and $h^{\mu\nu}$ are connected in first approximation by
$h^{\mu\nu}=-\phi^{\mu\nu}+\frac{1}{2}\eta^{\mu\nu}\phi^{\sigma}_{\  \sigma}$, the gauge transformation~(\ref{eq:2.13})
becomes the one of Eq.~(\ref{eq:2.11}), therefore the latter is the formulation of general covariance for
the Lorentz tensor field $h^{\mu\nu}(x)$, a property that was already recognized in~\cite{fi3} and discussed in 
a more general context in~\cite{wald}. From Eq.~(\ref{eq:2.12}) it follows also ${h'}^{\mu\nu}(x')=
\Lambda^{\mu}_{\;\rho}\Lambda^{\nu}_{\;\sigma}h^{\rho\sigma}(x)$ under a Lorentz transformation 
$\Lambda\in\mathcal{L}_{+}^{\uparrow}$.

The gauge condition relative to the gauge transformation~(\ref{eq:2.11}) is the Hilbert gauge condition
\begin{equation}
h^{\mu\nu}(x)_{,\nu}=0\, ,
\label{eq:2.14}
\end{equation}
because it remains invariant under~(\ref{eq:2.11}):
\begin{equation}
h^{\mu\nu}_{\: ,\nu}=0 \longrightarrow {h'}^{\mu\nu}_{\: ,\nu}= h^{\mu\nu}_{\: ,\nu}+\underbrace{\Box  f^{\mu}}_{=0} 
+\underbrace{f^{\nu}_{,\mu\nu}-f^{\gamma}_{,\mu\gamma}}_{=0} =0 \, .
\label{eq:2.15}
\end{equation}
By imposing the gauge condition~(\ref{eq:2.14}) on~(\ref{eq:2.10}), we arrive at the free wave equation
\begin{equation}
\square h^{\mu\nu}(x)=0\, ,
\label{eq:2.16}
\end{equation}
for the massless graviton field. 

Under the gauge transformation~(\ref{eq:2.11}), the linearized Riemann tensor
\begin{multline}
R^{\alpha\beta\mu\nu}_{\sss{L}}=\frac{\kappa}{2}\big[ h^{\alpha\mu}_{,\beta\nu}- h^{\beta\mu}_{,\alpha\nu}- 
                                                h^{\alpha\nu}_{,\beta\mu}+ h^{\beta\nu}_{,\alpha\mu}
              -\frac{1}{2}\eta^{\alpha\mu}h_{,\beta\nu}+\frac{1}{2}\eta^{\beta\mu}h_{,\alpha\nu}\\
              +\frac{1}{2}\eta^{\alpha\nu}h_{,\beta\mu}-\frac{1}{2}\eta^{\beta\nu}h_{,\alpha\mu}\big]\, ,
\end{multline}
the linearized Ricci tensor
\begin{equation}
R^{\mu\nu}_{\sss{L}}=\frac{\kappa}{2}\big[\Box h^{\mu\nu}-\frac{1}{2}\eta^{\mu\nu}\Box h - h^{\mu\rho}_{,\rho\nu}-
                   h^{\nu\rho}_{,\rho\mu}\big]\, ,
\end{equation}
and the linearized Ricci scalar
\begin{equation}
R_{\sss{L}}=\frac{\kappa}{2}\big[-\Box h -2 h^{\rho\sigma}_{,\rho\sigma}\big]\, ,
\end{equation}
remain invariant, whereas $\mathcal{L}_{\sss {HE}}^{\sss{(0)}}$ is changed by negligible  divergences:
\begin{equation}
{R'}^{\alpha\beta\mu\nu}_{\sss{}L}=R^{\alpha\beta\mu\nu}_{\sss{L}}\, ,\quad {R'}^{\mu\nu}_{\sss{L}}=R^{\mu\nu}_{\sss{L}}\, ,
\quad {R'}_{\sss{L}}=R_{\sss{L}}\, ,
\quad {\mathcal{L}'}_{\sss HE}^{\sss{(0)}}=\mathcal{L}_{\sss HE}^{\sss{(0)}}+\mathrm{divergences}\, .
\label{eq:2.17}
\end{equation}

\subsection{Quantization of Gravity}

We consider now the graviton field $h^{\mu\nu}(x)$ as a free quantum field which satisfies the wave 
equation~(\ref{eq:2.16}) and quantize it by imposing the following Lorentz covariant commutation rule:
\begin{equation}
\big[ h^{\alpha\beta}(x),h^{\mu\nu}(y) \big]=-i\, b^{\alpha\beta\mu\nu}\, D_{0}(x-y)\, ,
\label{eq:18}
\end{equation}
with
\begin{equation}
b^{\alpha\beta\mu\nu}:=\frac{1}{2}\left(  \eta^{\alpha\mu}\eta^{\beta\nu}+ 
  \eta^{\alpha\nu}\eta^{\beta\mu}-\eta^{\alpha\beta}\eta^{\mu\nu}\right)\, ,
\label{eq:18.1}
\end{equation}
and $D_{0}(x)$ is the mass-zero Jordan--Pauli causal distribution:
\begin{equation}
D_{0}(x)=\frac{1}{2\pi}\delta(x^2)\, \mathrm{sgn}(x^{0})=\frac{i}{(2\pi)^3}\int \!\!d^{4}p\, \delta (p^2)\,
     \mathrm{sgn}(p^{0})\, e^{-i\,p\cdot x} \, .
\end{equation}
The explicit free field representation for the graviton field $h^{\mu\nu}$ in~(\ref{eq:18}) will be given in Sec.~\ref{sec:freefield}.

Obviously, the ten components of $h^{\mu\nu}$ contain more than the true physical degrees of freedom of a massless
spin-2 field (see Sec.~\ref{sec:physsub}), this  additional freedom could be suppressed by a gauge 
condition $h^{\mu\nu}_{\: ,\nu}=0$ and a trace condition $h=0$. As in gauge theories these condition are 
disregarded at the beginning and considered later as conditions on the physical states.

In the causal approach the commutation rule~(\ref{eq:18}) is fixed by choosing as operator gauge transformation 
that of~(\ref{eq:2.11}) and by defining the gauge charge $Q$ (see Sec.~\ref{sec:gauge}), but it is independent
from the choice of the expansion variable in Eq.~(\ref{eq:gtilde}).

Alternatively, the form of the $b$-tensor~(\ref{eq:18.1}) can be traced 
back to the addition of the gauge fixing term~\cite{velt},~\cite{gu4}:
\begin{equation}
\mathcal{L}_{\sss{GF}}=\kappa^{-2}\big(\partial_{\nu}\tilde{g}^{\mu\nu}\big)\big(\partial_{\rho}\tilde{g}^{\mu\rho}\big)
=h^{\mu\nu}_{\: ,\nu} h^{\mu\rho}_{\: ,\rho}
\label{eq:19}
\end{equation}
to the free Lagrangian $\mathcal{L}_{\sss HE}$; this extension is necessary for the total Lagrangian to be invertible
(disregarding some divergences):
\begin{equation}
\mathcal{L}^{\sss (0)}_{\sss HE} + \mathcal{L}_{\sss{GF}}= \frac{1}{2}h^{\mu\nu}_{\: ,\rho}h^{\mu\nu}_{\: ,\rho}
                         -\frac{1}{4}h_{,\rho}h_{,\rho}=\frac{1}{2}h^{\mu\nu}_{\: ,\rho}b_{\mu\nu\alpha\beta}
                         h^{\alpha\beta}_{\: ,\rho}=-\frac{1}{2}h^{\mu\nu}b_{\mu\nu\alpha\beta}\Box h^{\alpha\beta}\, ,
\label{eq:21}
\end{equation}
so that we obtain the Feynman propagator
\begin{gather}
\langle\Omega| T\big\{h^{\alpha\beta}(x)h^{\mu\nu}(y)\big\}|\Omega \rangle =-i\, b^{\alpha\beta\mu\nu}\, 
                 D^{\sss{F}}_{0}(x-y)\, ,\nonumber \\
D^{\sss{F}}_{0}(x)=\frac{1}{(2\pi)^4}\int\!\! d^4 p\,  \frac{-1}{p^2 +i0}\, e^{-i\, p \cdot x}\, ,
\label{eq:feynman}
\end{gather}
where $|\Omega\rangle$ is the free Fock vacuum. 

The quantization of the gravitational field in the path-integral framework 
was carried out in~\cite{fade}, whereas the Feynman rules for the $S$-matrix were already given in~\cite{fratu}.

On the other side, the Feynman propagator in~(\ref{eq:feynman}) can be derived by means of the  commutation rule~(\ref{eq:18}), if we compute
the time-ordered product of the contraction between two field operators:
since the contraction is defined as
\begin{equation}
C\big\{ h^{\alpha\beta}(x) h^{\mu\nu}(y) \big\}:=\big[h^{\alpha\beta}(x)^{\sss (-)},h^{\mu\nu}(y)^{\sss (+)}\big]\, ,
\label{eq:22}
\end{equation}
where $h^{\mu\nu}(x)^{(\mp)}$ are the absorption and emission part of the free quantum field, respectively, we obtain
\begin{gather}
T\Big\{ C \big\{  h^{\alpha\beta}(x)h^{\mu\nu}(y) \big\}\Big\}=\Theta(x^0-y^0)\big[h^{\alpha\beta}(x)^{\sss (-)},
               h^{\mu\nu}(y)^{\sss (+)}\big]+ \nonumber \\
      +\Theta(y^0-x^0)\big[h^{\mu\nu}(y)^{\sss (-)},h^{\alpha\beta}(x)^{\sss (+)}\big] \nonumber \\
=-i\, b^{\alpha\beta\mu\nu}\,\big(\Theta(x^0-y^0)\,  D^{\sss (+)}_{0}(x-y) +\Theta(y^0-x^0)\, D^{\sss (+)}_{0}(y-x)\big)\nonumber\\
= -i\, b^{\alpha\beta\mu\nu}\,\big( D^{ret}_{0}(x-y)-D_{0}^{\sss (-)}(x-y)\big)
= -i\, b^{\alpha\beta\mu\nu}\, D^{\sss F}_{0}(x-y)\, .
\label{eq:23}
\end{gather}

The commutation rule~(\ref{eq:18}) leads also to the correct equal-time canonical commutation
relation between the field and its canonical conjugate momentum with respect to the gauge fixed Lagrangian
defined by
\begin{equation}
\Pi^{\mu\nu}(t,\bol{x})=\frac{\partial\big(\mathcal{L}^{\sss{(0)}}_{\sss HE} + \mathcal{L}_{\sss{GF}}\big)}
{\partial \dot{h}^{\mu\nu}}=\dot{h}^{\mu\nu}(t,\bol{x})-\frac{1}{2}\eta^{\mu\nu}\dot{h}(t,\bol{x})\, ,
\label{eq:24}
\end{equation}
so that we have
\begin{equation}
\big[h^{\alpha\beta}(t,\bol{x}),\Pi_{\mu\nu}(t,\bol{y})\big]=
+\frac{i}{2}\,\big\{\eta^{\alpha}_{\: \mu}\eta^{\beta}_{\: \nu}+\eta^{\alpha}_{\: \nu}\eta^{\beta}_{\: \mu}\big\}\,
\delta^{\sss{(3)}}(\bol{x}-\bol{y})=+i\, l^{\alpha\beta}_{\quad\mu\nu}\, \delta^{\sss{(3)}}(\bol{x}-\bol{y})\, ,
\label{eq:25}
\end{equation}
which agrees with the fundamental Poisson brackets derived in~\cite{baak} and~\cite{jose} and
where $l^{\alpha\beta\mu\nu}$ is the unity for rank-4 tensors.

\subsection{First Order Graviton Interaction}

Since the perturbative expansion for the $S$-matrix~(\ref{eq:1.1}) is in powers of the coupling constant $\kappa$,
we consider the normally ordered product
\begin{equation}
\begin{split}
T_{1}^{h}(x)& =i\, \kappa :\!{\mathcal{L}}_{\sss HE}^{ \sss{(1)}}(x)\!: \\
            & =i\,  \frac{\kappa}{2}  \big\{
                  + :\!h^{\rho\sigma}(x)h^{\alpha\beta}(x)_{,\rho}h^{\alpha\beta}(x)_{,\sigma}\!:
       -\frac{1}{2}:\!h^{\rho\sigma}(x)h(x)_{,\rho}h(x)_{,\sigma}\!:  \\
         &\quad\quad   +2:\!h^{\rho\sigma}(x)h^{\sigma\beta}(x)_{,\alpha}h^{\rho\alpha}(x)_{,\beta}\!:
                  +:\!h^{\rho\sigma}(x)h(x)_{,\alpha}h^{\rho\sigma}(x)_{,\alpha}\!:\\
      &\quad \quad     -2:\!h^{\rho\sigma}(x)h^{\alpha\rho}(x)_{,\beta}h^{\sigma\alpha}(x)_{,\beta}\!: \big\}\,,
\label{eq:26}
\end{split}
\end{equation}
as the  cubic interaction among gravitons or first order graviton  interaction. 
We will omit the double dots of the normal ordering and the space-time dependence, if the meaning is clear.

After quantization, Eq.~(\ref{eq:18}), the coupling~(\ref{eq:26}), completed by a suitable ghost-graviton
coupling term (see Sec.~\ref{sec:ghost}), can be used in perturbation theory to calculate quantum correction
to classical general relativity.

Two serious problems arise in this procedure. 
The first one is the non-renormalizability of quantum gravity due
to presence of two derivatives on the graviton fields in~(\ref{eq:26}) whose origin lies in the dimensionality
of the coupling constant: $[\kappa ]=\mathrm{mass}^{-1}$.
The second one is the non-polynomial character of $\mathcal{L}_{\sss HE}$, Eq.~(\ref{eq:2.7}), which reflects itself
into a `proliferation of couplings', \emph{i.e.} into an increasing polynomial degree in the interaction structure.

The first drawback, non-renormalizability, can be in some manner `cured' by means of the inductive causal 
construction of the $T_{n}$'s, Eq.~(\ref{eq:1.10}), which makes it possible to find finite and cutoff-free 
quantum corrections  for any process describable in the $S$-matrix framework, although the solution is
not quite clear with regard to physical predictability because of the increasing number of finite
normalization terms in the distribution splitting~(\ref{eq:1.8}) in each order of perturbation theory.

With regard to the second issues, we could try to generalize the preliminary result of~\cite{scho1} and the more
recent result of~\cite{well1}, which suggest that
the concept of `perturbative quantum operator gauge invariance' (see Sec.~\ref{sec:gauge}) may be able to explain the
higher polynomial couplings on a quantum level.

\section{Perturbative Quantum Operator Gauge Invariance}\label{sec:gauge}
\setcounter{equation}{0}

\subsection{Infinitesimal Asymptotic Gauge Transformations}

The classical gauge transformation~(\ref{eq:2.11}) can be implemented at the quantum level as follows
\begin{equation}
\begin{split}
h^{'\alpha\beta}(x)&=e^{-i\,\lambda\, Q}h^{\alpha\beta}(x)e^{i\,\lambda\, Q}\\
                   &=h^{\alpha\beta}(x)-i\,\lambda \big[Q,h^{\alpha\beta}(x)\big]
                           -\frac{\lambda^2}{2}\big[Q,\big[Q,h^{\alpha\beta}(x)\big]\big]+\cdots\, ,
\label{eq:27}
\end{split}
\end{equation}
with the Lorentz invariant and time-independent  gauge charge
\begin{equation}
Q:=\int\limits_{x^0 =const}\!\! d^{3}x\,\underbrace{ h^{\alpha\beta}(x)_{,\beta} {\stackrel{\longleftrightarrow}{ \partial_{x}^{0} }}
 u_{\alpha}(x)}_{=J^{0}(x)}\, ,
\label{eq:28}
\end{equation}
where the gauge current is conserved, $\partial_{\mu}^{x}J^{\mu}(x)=0$, if $\Box u^{\nu}(x)=0$.
The main feature of the gauge charge $Q$ lies in the fact that the graviton field appears here in the 
Hilbert gauge condition~(\ref{eq:2.14}).

In order to get a nilpotent ($Q^2 =0$) gauge charge, we have to quantize
the vector field $u^{\mu}(x)$, the ghost field, with its partner $\tilde{u}^{\nu}(x)$, the anti-ghost field 
(with $\Box \tilde{u}^{\nu}(x)=0$, too), as free fermionic vector fields through the anti-commutator
\begin{equation}
\big\{u^{\mu}(x),\tilde{u}^{\nu}(y)\big\}=i\, \eta^{\mu\nu}\, D_{0}(x-y) \, ,
\label{eq:29}
\end{equation}
whereas all other anti-commutators vanish.
The gauge charge $Q$ defines an infinitesimal gauge variation by
\begin{equation}
d_Q A:= QA-(-1)^{n_{\sss{G}} (A)} AQ\, ,
\label{eq:30}
\end{equation}
where $n_{\sss{G}} (A)$ is the number of ghost fields minus the number of anti-ghost fields in the Wick monomial $A$. The operator
$d_Q$ obeys also the Leibniz rule
\begin{equation}
d_Q (AB)=(d_Q A) B +(-1)^{n_{\sss{G}} (A)} A  d_Q B \, ,
\end{equation}
for arbitrary operators $A$ and  $B$.

The infinitesimal operator gauge variations of the fundamental asymptotic free quantum fields read:
\begin{gather}
d_Q h^{\alpha\beta}(x)  =\big[ Q, h^{\alpha\beta}(x)\big]=-i\, b^{\alpha\beta\rho\sigma} u^{\rho}(x)_{,\sigma}\, ,\nonumber\\
d_Q h^{\alpha\beta}(x)_{,\beta}  =\big[ Q, h^{\alpha\beta}(x)_{,\beta}\big]=-i\, b^{\alpha\beta\rho\sigma} 
               u^{\rho}(x)_{,\sigma\beta}=0\, ,\nonumber\\
d_Q u^{\alpha}(x) = \big\{ Q, u^{\alpha}(x)\big\}=0\, , \nonumber \\
d_Q \tilde{u}^{\alpha}(x) = \big\{ Q, \tilde{u}^{\alpha}(x)\big\}=i\, h^{\alpha\beta}(x)_{,\beta}\, .
\label{eq:31}
\end{gather}
The  $b$-tensor appearing in the quantization rule~(\ref{eq:18}) is consequently necessary for the operator
gauge transformation of  $h^{\alpha\beta}(x)$ to correspond to the classical one, Eq.~(\ref{eq:2.11}), after having
chosen the gauge charge $Q$.

\subsection{Perturbative Gauge Invariance and Ghost Coupling}\label{sec:ghost}

The asymptotic free fields are used in order to construct the time-ordered `renormalized' products $T_{n}$, Eq.~(\ref{eq:1.10}),
in the adiabatically switched $S$-matrix, Eq.~(\ref{eq:1.1}). Therefore, gauge invariance of the $S$-matrix can be directly
formulated as  conditions on the $T_{n}$'s.
Formally, we have $S$-matrix gauge invariance if
\begin{equation}
\lim_{g \to 1}\big( S'(g)-S(g)\big)=\lim_{g \to 1}\left(
-i\,\lambda \left[ Q,S(g)\right]+\text{higher commutators}\right)=0\, .
\label{eq:32}
\end{equation}
Then this  condition  becomes
\begin{equation}
\lim_{g \to 1}\big[ Q, S(g)\big]=0 \, .
\label{eq:33}
\end{equation}
Since the existence of the adiabatic limit in massless theories is problematic, we use the perturbative version
of the above equation.
Inserting in~(\ref{eq:33}) the perturbative expansion of $S(g)$, 
we see that the perturbative gauge condition for the $T_{n}$'s:
\begin{equation}
d_Q T_{n}(x_1 ,\ldots,x_n )=\text{divergence in the sense of vector analysis}\, ,
\label{eq:34}
\end{equation}
implies $S$-matrix gauge invariance, because divergences do not 
contribute in the adiabatic limit due to partial integration and 
Gauss' theorem. 

Already for $n=1$ the above requirement is not at all trivial, because for  QG we find that $d_Q T_{1}^{h}(x)\neq divergence $.
This requires the introduction of an interaction between gravitons and ghosts. We choose this first order ghost coupling
as in~\cite{kugo},~\cite{nishi} to be
\begin{equation}
\begin{split}
T_{1}^{u}=i\, \kappa\, \big(&+:\tilde{u}^{\nu}(x)_{,\mu} h^{\mu\nu}(x)_{,\rho} u^{\rho}(x):
                           -:\tilde{u}^{\nu}(x)_{,\mu}h^{\nu\rho}(x) u^{\mu}(x)_{,\rho}: \\
                         &- :\tilde{u}^{\nu}(x)_{,\mu}h^{\mu\rho}(x)u^{\nu}(x)_{,\rho}:
                           +:\tilde{u}^{\nu}(x)_{,\mu}h^{\mu\nu}(x)u^{\rho}(x)_{,\rho}: \big)\, ,
\label{eq:35}
\end{split}
\end{equation}
so that the sum of~(\ref{eq:26}) with~(\ref{eq:35}) preserves perturbative gauge invariance to first order as 
shown in~\cite{scho1}:
\begin{equation}
d_{Q}\big(T_{1}^{h}(x)+T_{1}^{u}(x)\big)=:
\partial_{\nu}^{x}T_{1/1}^{ \nu}(x)=\hbox{sum of  divergences}\, .
\label{eq:36}
\end{equation}
One form of $T_{1/1}^{\nu}(x)$, the so-called $Q$-vertex, was derived in~\cite{scho1}.
The ghost couplings in causal quantum gravity are analyzed in great detail in~\cite{scho2}.
The fermionic quantization of the ghost fields is not only necessary for having a nilpotent  $Q$,
but also for perturbative gauge invariance to be fulfilled.

The ghost fields, usually called Faddeev-Popov ghosts~\cite{fade}, are introduced
in the causal construction as a consequence of perturbative gauge invariance
for $n=1$ in  Eq.~(\ref{eq:34}). 
In the path-integral framework, the ghost fields appear as a consequence of 
the quantization after gauge fixing, but it was already noticed
by Feynman~\cite{fey1} that without ghost fields a unitarity breakdown occurs in second
order at the loop level.

Although the condition $d_{Q}T_{1}(x)=divergence$ seems to be rather easy to fulfil, it has two important
consequences.
First of all, it rules out the possibility of a renormalizable theory of quantum gravity~\cite{well1}, because for a 
renormalizable 
interaction $T_{1}(x)$, \emph{i.e.} without the two derivatives acting on the 
fields\footnote{with only one derivative it is
impossible to form a Lorentz scalar interaction term}, first order perturbative gauge invariance
entails  only the trivial solution $T_{1}(x)=0$.

The other interesting consequence pointed out  in~\cite{well1} is that the requirement  $d_{Q}T_{1}(x)=divergence$, where 
$T_{1}(x)$ is  the most general ansatz for the graviton coupling and  the most general ansatz for the ghost coupling,
selects a small number of possible theories and the Hilbert--Einstein graviton coupling $T_{1}^{\sss HE}$, 
Eq.~(\ref{eq:26}),
supplemented by the Kugo--Ojima ghost coupling $T_{1}^{\sss KO}$, Eq.~(\ref{eq:35}), lies among them, moreover all other 
couplings can be transformed in such a way that the most general coupling has the form
\begin{equation}
T_{1}(x)=T_{1}^{\sss HE}(x)+ T_{1}^{\sss KO}(x) + \text{divergence couplings}+d_{Q}(\tilde{u}hh +\tilde{u}\tilde{u}u)\, .
\end{equation}
The last term represents the so-called `coboundary terms' which, together with divergence terms, seem not to play
any physical r\^ ole.

The definition of the $Q$-vertex from Eq.~(\ref{eq:36}) allows us to give  a precise prescription on how the right side of
Eq.~(\ref{eq:34}) has to be inductively constructed.
We define the concept of `perturbative quantum operator gauge invariance' by the equation
\begin{equation}
d_{Q}T_{n}(x_{1},\ldots ,x_{n})=\sum_{l=1}^{n}\frac{\partial}{\partial x^{\nu}_{l}}\, 
T_{n/l}^{\nu}(x_{1},\ldots , x_{l},\ldots ,x_{n}) \, .
\label{eq:37}
\end{equation}
Here $T_{n/l}^{\nu}$ is the time-ordered renormalized product, obtained according to the inductive causal scheme, with a
$Q$-vertex at $x_{l}$, while all other $n-1$ vertices are ordinary $T_{1}$-vertices.

Analysis of the condition~(\ref{eq:37}) shows that perturbative gauge invariance can be spoiled by local 
terms, \emph{i.e.}
terms proportional to $:\!\mathcal{O}(x_{1},\ldots ,x_{n})\! : \delta^{\sss (4n-4)}(x_{1}-x_{n},\ldots ,x_{n-1}-x_{n})$, which
appear as a consequence of distribution splitting on both  sides  of Eq.~(\ref{eq:37}) in the tree graph sector.

If it is possible to absorb these local terms by suitable local normalization terms (see Eq.~(\ref{eq:1.10})) $N_{n}$ of
$T_{n}$ and $N_{n/l}^{\mu}$ of $T_{n/l}^{\mu}$ in such a way that the equation
\begin{equation}
d_{Q}\Big(T_{n}+N_{n}\Big)(x_{1},\ldots ,x_{n})=\sum_{l=1}^{n}\frac{\partial}{\partial x^{\nu}_{l}}\, 
\Big( T_{n/l}^{\nu}+N_{n/l}^{\nu} \Big) (x_{1},\ldots , x_{l},\ldots ,x_{n}) 
\label{eq:38}
\end{equation}
holds true, then we call the theory gauge invariant to $n$-th order.

The  analysis of perturbative gauge invariance  for $n=2$ was carried out in~\cite{scho1}, with the result that
the $N_{2}(x_{1},x_{2})$-terms, necessary in order for~(\ref{eq:38}) to hold, are local normalization terms of tree graphs 
with singular order $\omega\ge 0$ which agree exactly with the quartic expression in the expansion~(\ref{eq:2.7})
of the Hilbert--Einstein Lagrangian:
\begin{equation}
N_{2}(x_{1},x_{2})=i\, \kappa^2 :\!\mathcal{L}_{\sss HE}^{\sss{(2)}}(x_{1})\! : \delta^{\sss (4)} (x_{1}-x_{2})\ .
\end{equation}
In this way perturbative gauge invariance `generates' the four-graviton coupling and  the proliferation of couplings can 
be understood on a quantum level. 

Such a property was already observed in Yang--Mills theories~\cite{nonab},~\cite{ym1}: starting with an interaction 
among three gauge fields, perturbative gauge invariance to second order generates automatically the four gauge fields 
coupling and the
requirement of perturbative gauge invariance seems to be strong enough to select the correct Yang--Mills interaction
among all the possible ones.

\subsection{Finite Gauge Transformations}

The infinitesimal gauge transformations~(\ref{eq:31}) correspond to the finite gauge transformations~(\ref{eq:27}) 
generated
by $Q$. If we work out the latter we find
\begin{gather}
h^{'\mu\nu}(x)  =h^{\mu\nu}(x)-\lambda\, b^{\mu\nu\rho\sigma}u^{\rho}(x)_{,\sigma}+ \lambda^2\, Q\, h^{\mu\nu}(x)\, Q\, , \nonumber\\
u^{'\mu}(x)     =u^{\mu}(x)\big( 1+2\,i\,\lambda Q\big)\, , \nonumber  \\
\tilde{u}^{' \mu}(x)=\tilde{u}^{\mu}(x)+ \lambda \, h^{\mu\nu}(x)_{,\nu} +2\,i\,\lambda\, \tilde{u}^{\mu}(x)\, Q
+i\, \lambda^2\, h^{\mu\nu}(x)_{,\nu} Q\, . 
\label{eq:39}
\end{gather}
These equations and the corresponding infinitesimal version~(\ref{eq:31}) have some resemblance with the BRST 
transformations 
for the gravitational field. In fact, they can be obtained by restricting the latter to their linearized versions.

As we saw in Sec.~\ref{sec:ghost}, the lowest order graviton self-coupling alone is not gauge invariant. This is the reason
why ghost couplings must be introduced. This is in contrast to BRST theories of gravity~\cite{kugo},~\cite{nishi},
where the gravitational Lagrangian, on the one
hand, and the sum of the ghost interaction and the gauge fixing term, on the other hand are separately BRST invariant.

\section{Quantization of the Graviton Field in Fock Space}\label{sec:quant}
\setcounter{equation}{0}

\subsection{Classical Tensor Fields}

The classical tensor field $h^{\alpha\beta}(x)$, because of its transformation law $h^{'\mu\nu}(x^{'})=
\Lambda^{\mu}_{\;\rho}\Lambda^{\nu}_{\;\sigma}h^{\rho\sigma}(x)$ under the proper 
Lorentz group $\mathcal{L}_{+}^{\uparrow}$,
transforms as a symmetrized tensor product of two spinor representations $\mathcal{D}^{(1/2,1/2)}$.
The product of two such spinor representations can be decomposed into irreducible representations as follows
\begin{equation}
{\mathcal D}^{(1/2,1/2)} \otimes {\mathcal D}^{(1/2,1/2)} 
={\mathcal D}^{(1,1)}\oplus  {\mathcal D}^{(1,0)} \oplus  {\mathcal D}^{(0,1)}\oplus  {\mathcal D}^{(0,0)}\, .
\label{eq:40}
\end{equation}
The first representation ${\mathcal D}^{(1,1)}$ is nine-dimensional and given by symmetric traceless tensors. 
The second one
${\mathcal D}^{(1,0)}$ is three-dimensional and given by anti-self-dual tensors, while the third one ${\mathcal D}^{(0,1)}$
is also three-dimensional given by antisymmetric self-dual tensors. The last one ${\mathcal D}^{(0,0)}$ is the trivial
one-dimensional representation given by scalar fields. The dimensions add up correctly to $4\times 4=16$.

The classical symmetric tensor field $h^{\alpha\beta}$ has arbitrary trace, therefore has $9+1= 10$ 
independent components~\cite{wyss}
and transforms as the symmetrization of Eq.~(\ref{eq:40}), namely ${\mathcal D}^{(1,1)}\oplus  {\mathcal D}^{(0,0)}$.
On the other hand, a pure massless spin-2 field has $2\times 2+1 =5$ independent components, so that 5 subsidiary 
conditions
must be imposed~\cite{fi3},~\cite{siha}. Those can be chosen in the following Lorentz covariant way:
\begin{equation}
h^{\alpha\beta}(x)_{,\beta}=0\, \quad\text{and}\quad h^{\sigma}_{\:\sigma}=0\ .
\label{eq:41}
\end{equation}
The implementation of these equations on a quantum level will be the subject of Sec.~\ref{sec:physsub}.
The construction of the graviton field in the framework of the Bargmann-Wigner theory is presented 
in~\cite{nou1},~\cite{nou2} and~\cite{spe1}.

\subsection{The Quantum Fields \boldmath$H^{\alpha\beta}$ and \boldmath$\Phi$ and their Quantization}

Let us now consider a quantized graviton field. Eq.~(\ref{eq:40}) suggests us the following Lorentz covariant decomposition
of $h^{\alpha\beta}(x)$, according to ${\mathcal D}^{(1,1)}\oplus  {\mathcal D}^{(0,0)}$, into
\begin{equation}
h^{\alpha \beta}(x)=H^{\alpha\beta}(x) +\frac{1}{4}\,  \eta^{\alpha\beta}\, \Phi(x)\, ,
\label{eq:42}
\end{equation}
where $ H^{\alpha\beta}(x)$ represents a traceless symmetric tensor field defined as 
$H^{\alpha\beta}(x):=h^{\alpha\beta}(x)- \eta^{\alpha\beta}h(x)/4$ with $H^{\gamma}_{\ \gamma}=0$
(9 independent components) and  $ \Phi(x)$ a scalar field with $h^{\gamma}_{\ \gamma}=\Phi$.

Unlike the quantization proposed in~\cite{grigo}, where the scalar component $\Phi(x)$ is considered as a 
true ghost-like degree of freedom like  the ghost fields $u^{\mu}(x)$ and $\tilde{u}^{\mu}(x)$, we consider
this  scalar component $\Phi(x)$, necessary for the Lorentz covariant quantization given in~(\ref{eq:18}),
as a graviton component which turns out to be unphysical, see Sec.~\ref{sec:physsub}.

From Eq.~(\ref{eq:18}) we obtain the following commutation relations:
\begin{gather}
\big[ \Phi(x),\Phi(y)\big]=4\, i\, D_{0}(x-y)\, ,\nonumber \\
\big[ H^{\alpha\beta}(x),H^{\mu\nu}(y)\big]= -i\, t^{\alpha\beta\mu\nu}\, D_{0}(x-y)\, ,\nonumber \\
\big[ H^{\alpha\beta}(x),\Phi(y) \big]=0\, ,
\label{eq:43}
\end{gather}
with
\begin{equation}
t^{\alpha\beta\mu\nu}:=\frac{1}{2} \big(   \eta^{\alpha\mu} \eta^{\beta\nu} +
               \eta^{\alpha\nu} \eta^{\beta\mu}  - \frac{1}{2} \eta^{\alpha\beta} \eta^{\mu\nu} \big)\, .
\label{eq:44}
\end{equation}
With these new fields , the gauge transformation~(\ref{eq:2.11}) becomes
\begin{gather}
H^{\alpha\beta}(x) \longrightarrow H^{'\alpha\beta}(x)=H^{\alpha\beta}(x) +\lambda \,t^{\alpha\beta\gamma\delta} 
                                   u^{\gamma}(x)_{,\delta}\,, \nonumber \\
\Phi(x) \longrightarrow \Phi^{'}(x)= \Phi(x)-2\,\lambda \,u^{\sigma}(x)_{,\sigma} \, .
\label{eq:45}
\end{gather}
The Hilbert condition now reads $H^{\alpha\beta}(x)_{,\beta}+\Phi(x)^{,\alpha}/4=0$ and the first  operator
gauge transformation in~(\ref{eq:31}) now becomes
\begin{equation}
d_{Q} H^{\alpha\beta}(x)=-i\, t^{\alpha\beta\rho\sigma}u^{\rho}(x)_{,\sigma}\  \quad\text{and}
\quad d_{Q}\Phi(x)=+i\, u^{\sigma}(x)_{,\sigma}\, .
\label{eq:46}
\end{equation}
Only the linear combination $H^{\alpha\beta}(x)_{,\beta}+\Phi(x)^{,\alpha}/4$ has zero operator gauge
variation: $ d_{Q}\big(H^{\alpha\beta}(x)_{,\beta}+\Phi(x)^{,\alpha}/4\big)=0$.

\subsection{Free Field Representations for  \boldmath$H^{\alpha\beta}$ and \boldmath$\Phi$}\label{sec:freefield}

The Lorentz covariant quantizations~(\ref{eq:43}) have  to be verified by constructing explicit free field 
representations.

About free graviton field representation, there exists in the literature, among various proposals, a 
mainstream approach which is based on the Gupta--Bleuler method of indefinite metric Fock 
spaces~\cite{gu1},~\cite{gup3} and~\cite{mes3}.
But we choose to
avoid the  use of indefinite metric and realize all operators on a positive definite Fock space $\mathcal{F}$
which, however, later on will be given a Krein structure, see Sec.~\ref{sec:krein1}.

To be consistent with the Lorentz covariance of the commutation rules in Eq.~(\ref{eq:43}), some components 
of the $H$-field
and the $\Phi$-field must then be skew-Hermitian in contrast to others which are Hermitian.

We quantize provisionally all field components as being independent ones and choose for $H^{\alpha\beta}(x)$ and $\Phi(x)$
the following non-Lorentz covariant free field representations
\begin{gather}
\label{eq:47}
H^{\alpha\beta}(x)=(2\pi)^{-3/2}\int\!\! \frac{d^{3}k}{\sqrt{2\omega (\bol{k})}} \Big( A^{\alpha\beta}(\bol{k})e^{-i\,k \cdot x}
            +\eta^{\alpha\alpha}\eta^{\beta\beta} A^{\alpha\beta}(\bol{k})^{\dagger} e^{+i\,k\cdot x}\Big)\, ,\\
\Phi(x)=(2\pi)^{-3/2}\int\!\! \frac{d^{3}k}{\sqrt{2\omega (\bol{k})}}\Big( a(\bol{k})e^{-i\,k\cdot x}-
                      a(\bol{k})^{\dagger} e^{+i\,k\cdot x}\Big)\, ,
\label{eq:48}
\end{gather}
with $\omega(\bol{k})=|\bol{k}|$.
The absorption/creation operators satisfy the commutation relations 
\begin{equation}
\begin{split}
\big[ A^{\alpha\beta}(\bol{k}), A^{\mu\nu}(\bol{p})^{\dagger}\big]&=\eta^{\alpha\alpha}\eta^{\beta\beta}\,
                                                              t^{\alpha\beta\mu\nu}\,\delta^{\sss{(3)}}(\bol{k}-\bol{p}) \\
                                              &=  \tilde{t}^{\alpha\beta\mu\nu}\, \delta^{\sss{(3)}}(\bol{k}-\bol{p})\, ,      \\
\big[a(\bol{k}),a(\bol{p})^{\dagger}\big] &= 4\,\delta^{\sss{(3)}}(\bol{k}-\bol{p})\, .
\label{eq:50}
\end{split}
\end{equation}
Since we work in a fixed Lorentz system, the indices of $A^{\alpha\beta}$ and of $\tilde{t}^{\alpha\beta\mu\nu}$
lose their Lorentz character and label only a particular component of the $\tilde{t}$-tensor or of the operator $A^{\alpha\beta}$.

The factor $\eta^{\alpha\alpha}\eta^{\beta\beta}$ has the effect of changing the sign of $A^{\alpha\beta}(\bol{k})^{\dagger}$
in the $0i$-components, $i=1,2,3$: $H^{0i}\sim (A^{0i} - {A^{0i}}^{\dagger})$ and analogously for the $\Phi$-field:
$\Phi\sim (a-a^{\dagger})$. Therefore these field components are skew-adjoint:
\begin{equation}
{H^{0i}}^{\dagger}=-H^{0i}\quad\text{and}\quad \Phi^{\dagger}=-\Phi\, ,
\label{eq:51}
\end{equation}
so that the commutation rules~(\ref{eq:43}) are consistent with the free field definitions~(\ref{eq:47}) and~(\ref{eq:48}).

This is not a serious drawback because it will turn out (see Sec.~\ref{sec:physsub}) that these components, 
which spoil self-adjointness, are unphysical and  their  expectation values between  physical states  vanish.

The advantage of choosing the `wrong' sign and breaking self-adjointness as well as explicit Lorentz covariance lies in the
fact that the commutation rules for the $0i$-components
\begin{equation}
\big[ A^{0i}(\bol{k}), A^{0j}(\bol{p})^{\dagger}\big]=\tilde{t}^{\,0i0j}\, \delta^{\sss{(3)}}(\bol{k}-\bol{p})
               =\frac{1}{2}\,\delta^{ij}\,\delta^{\sss{(3)}}(\bol{k}-\bol{p})\, ,
\label{eq:52}
\end{equation}
and for the scalar field in Eq.~(\ref{eq:50}) have a positive right side.

The $\tilde{t}^{\alpha\beta\mu\nu}$-tensor has the following values:
\begin{center}
\begin{tabular}{|l|l|l|l|l|l|l|l|} 
\hline
$\tilde{t}^{(\alpha\beta , \mu\nu)}$ & $(00,00)$ & $(00,ii)$ & $(0i,0i)$ & $(ii,ii)$ & $(ii,jj)$ & $(ij,ij)$ &otherwise  \\   
\hline
value & \ \  3/4  &\ \  1/4  &\ \  1/2  &  \ \  3/4  & \ \  -1/4   &\ \    1/2 & \ \ \ \ \ 0 \\
\hline
\end{tabular}
\end{center}
with $i, j =1, 2, 3; i \ne j$.
From this table we see that the $\tilde{t}^{\alpha\beta\mu\nu}$-tensor is neither diagonal: 
$\tilde{t}^{\alpha\beta\mu\nu}=0$ for $(\alpha\beta)\neq (\mu\nu)$, nor positive definite: $\tilde{t}^{\alpha\beta\mu\nu}>0,\ \forall
\alpha,\beta,\mu,\nu$, although it is positive for the diagonal terms: $\tilde{t}^{\alpha\beta\alpha\beta}>0$ for
$(\alpha\beta)=00, 0i, ii, ij$. Moreover, until now we have not been making use of the fact that $H^{\sigma}_{\ \sigma}=0$,
which certainly has implications on the $A^{\alpha\beta}$-operators.

In order to remedy these  defects, we define new absorption operators:
\begin{equation}
\begin{split}
A^{00} &= \frac{1}{2}(+ {a}^{11} + {a}^{22}   + {a}^{33} )\, , \\
A^{11} &= \frac{1}{2}(- {a}^{11} + {a}^{22}   + {a}^{33} )\, , \\
A^{22} &= \frac{1}{2}(+ {a}^{11} - {a}^{22}   + {a}^{33} )\, , \\
A^{33} &= \frac{1}{2}(+ {a}^{11} + {a}^{22}   - {a}^{33} )\, ;
\end{split}
\label{eq:53}
\end{equation}
and analogously for the creation operators. Then we  obtain the commutation
relations
\begin{equation}
\big[ {a}^{ii}(\bol{k}),{a}^{jj}(\bol{p})^{\dagger}\big]=\delta^{ij}\,  \delta^{\sss{(3)}}(\bol{k}-\bol{p})\, ,
\label{eq:54}
\end{equation}
which have also positive right side.

Note that the operators ${a}^{00}$ and ${a}^{{00}^{\dagger}}$ do not appear here because this operator pair
is superfluous due to the trace condition $H^{\gamma}_{\ \gamma}=0$: from the definitions~(\ref{eq:53}) we get
\begin{equation}
\sum_{j=1}^{3} A^{jj}=A^{00}\, .
\label{eq:55}
\end{equation}
The equations~(\ref{eq:53}) can be solved for the $a^{ii}$-operators:
\begin{equation}
a^{11}=A^{00}-A^{11}\, , \quad a^{22}=A^{00}-A^{22}\, , \quad  a^{33}=A^{00}-A^{33}\, .
\label{eq:56}
\end{equation}
This means that the $A^{\alpha\alpha}$-operators are not the fundamental ones and must be replaced by the $a^{ii}$'s in the construction
of the Fock space.

\subsection{Fock Space Construction}\label{sec:fock}

We  quantize the $H^{\alpha\beta}$-graviton and the scalar graviton $\Phi$ as
10 independent scalar fields. With the following  definitions for the absorption operators
\begin{align}
B^{0i}(\bol{k})&:= A^{0i}(\bol{k})\, ,  &            B^{ij}(\bol{k})&:=A^{ij}(\bol{k})\, , \\
B^{ii}(\bol{k})&:= a^{ii}(\bol{k})\, ,  &            B^{00}(\bol{k})&:=a(\bol{k})/2 \, ;
\label{eq:57}
\end{align}
and analogous ones for the creation operators, we can recast the commutation rules~(\ref{eq:50}),~(\ref{eq:52}) and~(\ref{eq:54})
in the form
\begin{equation}
\big[ B^{\alpha\beta}(\bol{k}), B^{\mu\nu}(\bol{p})^{\dagger}\big]=  \tilde{l}^{\alpha\beta\mu\nu}\, 
\delta^{\sss{(3)}}(\bol{k}-\bol{p})\, ,
\label{eq:58}
\end{equation}
with
\begin{equation}
\tilde{l}^{\alpha\beta\mu\nu}=\frac{1}{2}\big(   \delta ^{\alpha\mu} \delta^{\beta\nu} +\delta^{\alpha\nu} \delta^{\beta\mu}\big)\, .
\end{equation}
The $\tilde{l}$-tensor is diagonal and always positive
\begin{center}
\begin{tabular}{|l|l|l|l|l|l|} 
\hline
$\tilde{l}^{(\alpha\beta , \mu\nu)}$ & $(00,00)$ & $(0i,0i)$ & $(ii,ii)$  & $(ij,ij)$ & otherwise  \\   
\hline
value & \ \ \quad 1 &\ \  1/2  &\ \  \quad  1  &  \ \  1/2   & \ \ \ \ \ 0 \\
\hline
\end{tabular}
\end{center}
with $i, j =1, 2, 3; i \ne j$.

Let us consider the multi-particle Hilbert space $\mathcal{H}^{n_{00}, n_{01},\ldots ,n_{23}, n_{33}}$ with $n_{00}$ $B^{00}$-particles,
$n_{01}$ $B^{01}$-particles,\ldots , $n_{23}$ $B^{23}$-particles and $n_{33}$ $B^{33}$-particles, then the graviton Fock space 
is defined by
\begin{equation}
\mathcal{F}:=\bigoplus_{n_{00},n_{01},\ldots ,n_{23}, n_{33}=0}^{\infty} \mathcal{H}^{n_{00}, n_{01},\ldots ,n_{23},n_{33}}\, .
\label{eq:59}
\end{equation}
One-graviton states can be described by true Lorentz tensor potentials, or `wave-functions', 
$\va^{\alpha\beta}(x)$. For simplicity
we do not write explicitly the time-dependence which is of the form $\exp(i\, \omega(\bol{k})\, t)$. Then
one-graviton states are linear combinations  of the form
\begin{gather}
|\Phi\rangle =B^{\dagger}(\va)\,|\Omega\rangle=\int\!\!d^{3}k\sum_{\alpha,\beta=0}^{3}\hat{\va}^{\alpha\beta}(\bol{k})\, 
B^{\alpha\beta}(\bol{k})^{\dagger}\,|\Omega\rangle\, ,\nonumber \\
|\Phi\rangle \in\Big( \mathcal{H}^{1,0,\ldots,0}\oplus\cdots\oplus \mathcal{H}^{0,0,\ldots,0,1}\Big) \, ,
\label{eq:60}
\end{gather}
where $|\Omega\rangle$ is the free Fock vacuum.

The scalar product $\langle\Psi|\Phi\rangle$ is evaluated by means of Eq.~(\ref{eq:58})
\begin{equation}
\langle\Psi|\Phi\rangle=\sum_{\alpha,\beta=0}^{3}\int\!\! d^{3}k\, \hat{\psi}^{\alpha\beta}(\bol{k})^{\ast}
\hat{\va}^{\alpha\beta}(\bol{k}) =:(\psi,\va)_{\sss{E}}
\label{eq:61}
\end{equation}
and the norm obtained from this scalar product is positive definite. The subscript `$E$' informs us that the sum is a Euclidean one and
not a Minkowski one.

Generalization to $n$-graviton states is straightforward
\begin{equation}
\begin{split}
|\Phi_{\sss{(n)}}\rangle & =\frac{1}{\sqrt{n!}}\, B^{\dagger}(\va_{1})\ldots B^{\dagger}(\va_{n})\, |\Omega\rangle   \\
                   & =\frac{1}{\sqrt{n!}}\,\int\!\! d^{3}p_{1}\ldots\int\!\! d^{3}p_{n} \sum_{\mu_{1},\nu_{1}=0}^{3}\ldots
\sum_{\mu_{n},\nu_{n}=0}^{3}\underbrace{\hat{\va}^{\mu_{1}\nu_{1}}_{1}(\bol{p}_{1})\ldots\hat{\va}^{\mu_{n}\nu_{n}}_{n}(\bol{p}_{n})}_
{=\hat{\Phi}_{(n)}^{\mu_{1}\nu_{1},\ldots,\mu_{n}\nu_{n}}(\bol{p}_{1},\ldots,\bol{p}_{n})}\cdot\\
&\quad\qquad\quad\cdot B^{\mu_{1}\nu_{1}}(\bol{p}_{1})^{\dagger}\ldots B^{\mu_{n}\nu_{n}}(\bol{p}_{n})^{\dagger}\, |\Omega\rangle \ ,
\label{eq:62}
\end{split}
\end{equation}
and
\begin{equation}
|\Phi_{\sss{(n)}}\rangle\in \mathcal{H}^{\sss{(n)}}=\bigoplus_{n_{00}+\cdots +n_{33}=n} 
\mathcal{H}^{n_{00}, n_{01},\ldots ,n_{23}, n_{33}} \ .
\label{eq:63}
\end{equation}
The generalized tensor potential  ${\Phi}_{\sss{(n)}}^{\mu_{1}\nu_{1},\ldots,\mu_{n}\nu_{n}}(x_1 ,\ldots, x_n)$ 
is totally symmetric
\begin{equation}
\hat{\Phi}_{\sss (n)}^{\mu_{1}\nu_{1},\ldots,\mu_{n}\nu_{n}}(\bol{p}_{1},\ldots,\bol{p}_{n})=
\hat{\Phi}_{\sss (n)}^{\mu_{\pi(1)}\nu_{\pi(1)},\ldots,\mu_{\pi(n)}\nu_{\pi(n)}}(\bol{p}_{\pi(1)},\ldots,\bol{p}_{\pi(n)})\ ,
\label{eq:64}
\end{equation}
under $\pi\in {\sigma}_{n}$ and transforms as
\begin{multline}
{\Phi'}_{\sss (n)}^{\mu_{1}\nu_{1},\ldots,\mu_{n}\nu_{n}}(x'_1 ,\ldots, x'_n)=
\big( \Lambda^{\mu_1}_{\ \alpha_1} \Lambda^{\nu_1}_{\ \beta_1}\big)\ldots
\big( \Lambda^{\mu_n}_{\ \alpha_n} \Lambda^{\nu_n}_{\ \beta_n}\big)\cdot\, \\
\cdot{\Phi}_{\sss (n)}^{\alpha_{1}\beta_{1},\ldots,\alpha_{n}\beta_{n}}(\Lambda^{-1}(x_1 -a) ,\ldots, 
\Lambda^{-1}(x_n -a))\, ,
\label{eq:65}
\end{multline}
under Poincar\'e transformations $(a,\Lambda)\in\mathcal{P}_{+}^{\uparrow}$.

The definition of the absorption operator reads
\begin{multline}
\Big( B(\va)\, \hat{\Phi}\Big)_{\sss (n)}^{\mu_{1}\nu_{1},\ldots,\mu_{n}\nu_{n}}
(\bol{p}_{1},\ldots,\bol{p}_{n})=\\ =\sqrt{n+1}\, \int\!\! d^{3}p \sum_{\rho,\sigma=0}^{3}\hat{\va}^{\rho\sigma}
(\bol{p})^{\ast}\,\tilde{l}^{\rho\sigma\mu_{0} \nu_{0}}\,
\hat{\Phi}_{\sss (n+1)}^{\mu_{0} \nu_{0},\mu_{1}\nu_{1},\ldots,\mu_{n}\nu_{n}}
(\bol{p}, \bol{p}_{1},\ldots,\bol{p}_{n})\\
=\sqrt{n+1}\, \int\!\! d^{3}p \sum_{\rho,\sigma=0}^{3}\hat{\va}^{\rho\sigma}(\bol{p})^{\ast}\, \hat{\Phi}_{\sss (n+1)}^
{\rho\sigma,\mu_{1}\nu_{1},\ldots,\mu_{n}\nu_{n}} (\bol{p}, \bol{p}_{1},\ldots,\bol{p}_{n})\, ,
\label{eq:66}
\end{multline}
and for the creation operator
\begin{multline}
\Big( B(\va)^{\dagger}\, \hat{\Phi}\Big)_{\sss (n)}^{\mu_{1}\nu_{1},\ldots,\mu_{n}\nu_{n}}
(\bol{p}_{1},\ldots,\bol{p}_{n})=\\
= \frac{1}{\sqrt{n}}\sum_{i=1}^{n} \hat{\va}^{\mu_{i}\nu_{i}}(\bol{p}_{i})\,
\hat{\Phi}_{\sss (n-1)}^{\mu_{1} \nu_{1},\ldots , \, \slash\!\!\!\!{\mu}_{i}\slash\!\!\! {\nu}_{i},\ldots,\mu_{n}\nu_{n}}
(\bol{p}_{1},\ldots, \slash\!\!\!\!{\bol{p}}_{i},\ldots,\bol{p}_{n})\\
=\sqrt{n}\, S_{n}^+ \big(\hat{\va}\otimes\hat{\Phi}_{\sss (n-1)}\big)^{\mu_{1}\nu_{1},\ldots,\mu_{n}\nu_{n}}
(\bol{p}_{1},\ldots,\bol{p}_{n}) \, ,
\label{eq:67}
\end{multline}
so that the commutation rule is
\begin{equation}
\Big(\big[ B(\psi), B(\va)^{\dagger}\big]\,\hat{\Phi}\Big)_{\sss (n)}^{\mu_{1}\nu_{1},\ldots,\mu_{n}\nu_{n}}
(\bol{p}_{1},\ldots,\bol{p}_{n})=(\psi,\va)_{\sss E}\,\hat{\Phi}_{\sss (n)}^{\mu_{1}\nu_{1},\ldots,\mu_{n}\nu_{n}}
(\bol{p}_{1},\ldots,\bol{p}_{n})\, ,
\label{eq:68}
\end{equation}
and the Fock space scalar product reads
\begin{equation}
\begin{split}
\langle\Phi |\Psi \rangle=\Phi_{\sss (0)}^{\ast}\Psi_{\sss (0)}+\sum_{n=1}^{\infty}\Big(\prod_{i=1}^{n}\int\!\!d^{3}p_i 
\sum_{\mu_i , \nu_i =0}^{3}\Big)
&\,  \hat{\Phi}_{\sss (n)}^{\mu_{1}\nu_{1},\ldots,\mu_{n}\nu_{n}}(\bol{p}_{1},\ldots,\bol{p}_{n})^{\ast}\cdot \\
&\cdot\hat{\Psi}_{\sss (n)}^{\mu_{1}\nu_{1},\ldots,\mu_{n}\nu_{n}}(\bol{p}_{1},\ldots,\bol{p}_{n})\, .
\label{eq:69}
\end{split}
\end{equation}

\section{Physical Subspace Characterization}\label{sec:physsub}
\setcounter{equation}{0}

According to the general theory~\cite{stro}, if the graviton field $h^{\mu\nu}$ has to be  a Poincar\'e invariant  
operator-valued  distribution (see Sec.~\ref{sec:poinc}), it cannot be  a pure spin-2 field, therefore we cannot 
have a local theory involving only physical gravitons. States corresponding to spin-1 and
spin-0 must also  be present in the Fock  space and  auxiliary conditions like Eq.~(\ref{eq:41}) must be imposed 
to single out the physical states~\cite{gu1},~\cite{gup3} and~\cite{mes3}.
 
The Fock space for the graviton previously constructed is large: we had to quantize the ten independent components of
the $H^{\alpha\beta}$-field and of the $\Phi$-field.

Since the true physical states of a spin-2 massless particle are represented by only two $\pm2$-helicity states, 
we have to reduce the number of independent components from ten to two.
This reduction implies a selection of a subspace $\mathcal{F}_{phys}$, the `physical subspace', in $\mathcal{F}$ 
that contains only these `physical' particle states.

This task is accomplished with the help of the gauge charge $Q$, Eq.~(\ref{eq:28}), and we can specify the physical 
subspace through the definition
\begin{equation}
\mathcal{F}_{phys}:=\ker\Big( \big\{ Q, Q^{\dagger} \big\} \Big)\, .
\label{eq:70}
\end{equation}
The motivation  for such a definition of $\mathcal{F}_{phys}$ resides in the fact that the operator
$\big\{ Q, Q^{\dagger} \big\}$ has the form of a sum of particle number operators which `counts' the 
unphysical degrees of freedom, therefore the kernel of $\big\{ Q, Q^{\dagger} \big\}$ does not contain them.

\subsection{Ghost Fields Quantization}

Before embarking in the calculation of $\big\{ Q, Q^{\dagger} \big\}$, we need also to quantize the ghost and anti-ghost 
vector fields.

They satisfy the Lorentz covariant anti-commutation rule~(\ref{eq:29}) and we assume that creation/absorption operators
satisfy the anti-commutation relations
\begin{gather}
\big\{ c^{\mu}(\bol{p}),c^{\nu}(\bol{k})^{\dagger}\big\} = \delta^{\mu\nu}\,  \delta^{\sss (3)}
(\bol{p} - \bol{k})\, ,\nonumber \\
\big\{ b^{\mu}(\bol{p}),b^{\nu}(\bol{k})^{\dagger}\big\} = \delta^{\mu\nu}\,  \delta^{\sss (3)}
(\bol{p} - \bol{k})\, ,
\label{eq:71}
\end{gather}
whereas all other anti-commutators vanish.

The Fock space for the ghost fields $\mathcal{F}_{\sss G}$ can be constructed starting with~(\ref{eq:71}) as usual in the
framework of second quantization for fermionic particles.

The ghost free field representations are then
\begin{equation}
\begin{split}
u^{\nu}(x)&=(2\pi)^{-3/2}\int\!\!\frac{d^{3}p}{\sqrt{2\omega(\bol{p})}} \Big( +b^{\nu}(\bol{p})e^{-i\,p\cdot x}
-\eta^{\nu\nu}{c^{\nu}(\bol{p})}^{\dagger} e^{i\,p\cdot x} \Big)\, ,\\
\tilde{u}^{\nu}(x)&=(2\pi)^{-3/2}\int\!\!\frac{d^{3}p}{\sqrt{2\omega(\bol{p})}} \Big( -c^{\nu}(\bol{p})e^{-i\,p\cdot x}
-\eta^{\nu\nu}{b^{\nu}(\bol{p})}^{\dagger} e^{i\,p\cdot x} \Big)\, .
\label{eq:72}
\end{split}
\end{equation}
Other issues about the ghost fields will be discussed in Sec.~\ref{sec:krein2}.

\subsection{The Physical Subspace $\mathcal{F}_{phys}$}

We must be careful and discriminate between the true Lorentz indices of $k^{\alpha}$ and the `labelling' 
indices of the operators $A^{\alpha\beta}$, $c^{\mu}$ and $b^{\nu}$. These latter will be always written as superscript.

First we compute the gauge charge $Q$ from~(\ref{eq:28})
\begin{equation}
Q=\int\limits_{x^0 =const}\!\! d^{3}x\,\Big( H^{\alpha\beta}(x)_{,\beta}+\frac{1}{4}\Phi(x)^{,\alpha}\Big)
 {\stackrel{\longleftrightarrow}{ \partial_{x}^{0} }} u^{\gamma}(x)\, \eta_{\alpha\gamma} \, ,
\end{equation}
in momentum space
\begin{equation}
Q=\int\!\!d^{3}k\,\Big( A^{\alpha}(\bol{k})^{\dagger} b^{\gamma}(\bol{k}) -B^{\alpha}(\bol{k}) c^{\gamma}(\bol{k})^{\dagger} \Big) 
\eta_{\alpha\gamma}\, ,
\label{eq:73}
\end{equation}
where
\begin{gather}
A^{\alpha}(\bol{k}):=\eta^{\alpha\alpha}\eta^{\beta\beta} k_{\beta} A^{\alpha\beta}(\bol{k}) -\frac{k^{\alpha}}{4}
                 a(\bol{k}) \, ,\nonumber \\
B^{\alpha}(\bol{k}):=\eta^{\alpha\alpha} k_{\beta} A^{\alpha\beta}(\bol{k}) +\eta^{\alpha\alpha} \frac{k^{\alpha}}{4} a(\bol{k}) \,  ;
\label{eq:74}
\end{gather}
so that 
\begin{equation}
\begin{split}
\big\{ Q, Q^{\dagger} \big\}=\int\!\!d^{3}k\int\!\! d^{3}p     \Big[ 
&+A^{\alpha}(\bol{k})^{\dagger} A^{\mu}(\bol{p})\, \big\{ b^{\gamma}(\bol{k}), b^{\nu}(\bol{p})^{\dagger}\big\} \\
&+B^{\mu}(\bol{p})^{\dagger} B^{\alpha}(\bol{k})\, \big\{ c^{\nu}(\bol{p}), c^{\gamma}(\bol{k})^{\dagger}\big\}\\
&+b^{\nu}(\bol{p})^{\dagger} b^{\gamma}(\bol{k})\, \big[  A^{\mu}(\bol{p}), A^{\alpha}(\bol{k})^{\dagger} \big]\\
&+c^{\gamma}(\bol{k})^{\dagger} c^{\nu}(\bol{p})\, \big[  B^{\alpha}(\bol{k}), B^{\mu}(\bol{p})^{\dagger} \big] \Big] 
\eta_{\alpha\gamma}\eta_{\mu\nu} \, .
\label{eq:75}
\end{split}
\end{equation}
Using~(\ref{eq:50}),~(\ref{eq:71}) and the definitions~(\ref{eq:74}) we obtain
\begin{equation}
\begin{split}
\big\{ Q, Q^{\dagger} \big\}=\int\!\!d^{3}k \Bigg[&+ \sum_{\alpha=0}^{3}\Big( A^{\alpha}(\bol{k})^{\dagger} A^{\alpha}(\bol{k}) +
B^{\alpha}(\bol{k})^{\dagger} B^{\alpha}(\bol{k}) \Big) \\
&+\omega^2 (\bol{k}) \sum_{\alpha=0}^{3} \Big( b^{\alpha}(\bol{k})^{\dagger}b^{\alpha}(\bol{k})
+c^{\alpha}(\bol{k})^{\dagger}c^{\alpha}(\bol{k}) \Big) \\
&+\frac{1}{2} \Big(\sum_{\alpha=0}^{3} k_{\alpha}c^{\alpha}(\bol{k})^{\dagger}\Big)
           \Big(\sum_{\beta=0}^{3} k_{\beta}c^{\beta}(\bol{k})\Big) \\
&+\frac{1}{2} \Big(\sum_{\alpha=0}^{3} k^{\alpha}b^{\alpha}(\bol{k})^{\dagger}\Big)
           \Big(\sum_{\beta=0}^{3} k^{\beta}b^{\beta}(\bol{k})\Big)   \Bigg]\, .
\label{eq:76}
\end{split}
\end{equation} 
In the ghost sector of $\big\{ Q, Q^{\dagger} \big\}$, all ghost degrees of freedom appear, therefore in 
$\ker \big( \big\{ Q, Q^{\dagger} \big\}\big)$ no ghost states are allowed.

The graviton sector of $\big\{ Q, Q^{\dagger} \big\}$ seems to consist of a sum of eight particle number operators, 
but we have to work it out a little bit more to see it clearly.
Substituting back
\begin{equation}
\begin{split}
A^{0}&=\omega \Big(+ A^{00}-A^{0}_{\parallel} -\frac{1}{4} a \Big) \, , \\
A^{i}&=\omega \Big(- A^{0i}+A^{i}_{\parallel} -\frac{k^{i}}{4\omega} a \Big) \, , \\
B^{0}&=\omega \Big(+ A^{00}+A^{0}_{\parallel} +\frac{1}{4} a \Big) \, , \\
B^{i}&=\omega \Big(- A^{0i}-A^{i}_{\parallel} -\frac{k^{i}}{4\omega} a \Big) \, ;
\label{eq:77}
\end{split}
\end{equation}
where
\begin{equation}
A^{\mu}_{\parallel}(\bol{k}):=\frac{k_{i}}{\omega (\bol{k})}\,A^{\mu i}(\bol{k})
\label{eq:78}
\end{equation}
is the absorption operator for the $\mu$-longitudinal mode, in Eq.~(\ref{eq:76}) we obtain for the graviton sector of
$\big\{ Q, Q^{\dagger} \big\}$
\begin{multline}
\big\{ Q, Q^{\dagger} \big\} \Big|_{\substack{ \text{graviton} \atop \text{sector}}}=
\int\!\!d^{3}k\,  2\,\omega^2 (\bol{k})\, \Bigg[+ \sum_{\mu=0}^{3}\Big( A^{0 \mu}(\bol{k})^{\dagger} A^{0 \mu}(\bol{k}) +
A^{\mu}_{\parallel}(\bol{k})^{\dagger} A^{\mu}_{\parallel}(\bol{k}) \Big) \\ +\frac{1}{8} a(\bol{k})^{\dagger}a(\bol{k}) \Bigg] \, .
\label{eq:79}
\end{multline}
Apparently there is an over-counting: we have  four $0\mu$- and four $\mu$-longitudinal modes, as well as the 
scalar component $a$, but $A^{0}_{\parallel}$ is not independent, being a linear combination of the 
$A^{0i}$-operators and we have not taken into account that $\eta_{\alpha\beta}A^{\alpha\beta}=0$.

For this purpose let us choose a reference frame in which $k^{\mu}=(\omega,0,0,\omega)$ is parallel to the 
third axis, because obviously the unphysical graviton modes depend on $\bol{k}$, and substitute the $A^{\mu\mu}$'s 
by the $a^{ii}$'s, Eq.~(\ref{eq:53}), so that the integrand of $\big\{ Q, Q^{\dagger} \big\}$ in Eq.~(\ref{eq:79}) becomes
\begin{multline}
2\, \omega^2 \, \big[ +2\, {A^{03}}^{\dagger} A^{03} + {A^{01}}^{\dagger} A^{01} + {A^{02}}^{\dagger} A^{02} +{A^{13}}^{\dagger} A^{13}+
                      {A^{23}}^{\dagger} A^{23} \big] +\\
+\omega^2 \, \big[ + {a^{11}}^{\dagger}a^{11}+ {a^{22}}^{\dagger}a^{22}+
 {a^{33}}^{\dagger}a^{33} + {a^{11}}^{\dagger}a^{22}  + {a^{22}}^{\dagger}a^{11} + \frac{1}{4} a^{\dagger}a    \big]\, .
\label{eq:80}
\end{multline}
With the definitions
\begin{gather}
J_{\pm}(\bol{k}):=\frac{ {a}^{11}(\bol{k}) \pm {a}^{22}(\bol{k})}{ \sqrt{2}}\, , \nonumber \\ 
\big[ J_{\pm}(\bol{k}), J_{\pm}(\bol{p})^{\dagger}\big]=\delta^{\sss (3)}(\bol{k}-\bol{p})\, , \quad 
\big[ J_{\pm}(\bol{k}), J_{\mp}(\bol{p})^{\dagger}\big]=0 \, ,
\label{eq:81}
\end{gather}
we find that the integrand of $\big\{ Q, Q^{\dagger} \big\}$ now reads 
\begin{multline}
2\, \omega^2 \, \big[ + {A^{01}}^{\dagger} A^{01} + {A^{02}}^{\dagger} A^{02} + 2\, {A^{03}}^{\dagger} A^{03}  
+{A^{13}}^{\dagger} A^{13}+\\
 +  {A^{23}}^{\dagger} A^{23} +  \frac{1}{8} a^{\dagger}a 
+\frac{1}{2}{a^{33}}^{\dagger} a^{33} +J_{+}^{\dagger}J_{+} \big]\, ,
\label{eq:82}
\end{multline}
which is manifestly the sum of  particle number operators for unphysical modes of the graviton field in the 
chosen reference frame: the two remaining physical modes for fixed $\bol{k}$ are created from the 
Fock vacuum $|\Omega\rangle$ by $J_{-}(\bol{k})^{\dagger}$ and $A^{12}(\bol{k})^{\dagger}$ in close analogy 
to the classical reduction of the degrees of freedom in a plane gravitational tensor wave 
$h^{\alpha\beta}_{cl.}(x)=\varepsilon_{cl.}^{\alpha\beta}(\bol{k})e^{-i\,k\cdot x}$
with polarization tensor $\varepsilon_{cl.}^{\alpha\beta}(\bol{k})$. Therefore Eq.~(\ref{eq:70}) defines in 
a correct manner the physical subspace.  

In fact, the physical modes can be described by two real polarization tensors
\begin{equation}
\varepsilon^{\alpha\beta}_{1}=\frac{1}{\sqrt{2}}\left( \begin{array}{cccc}
0 & 0  &  0  &  0   \\
0 & 1  &  0  &  0   \\
0 & 0  & -1  &  0  \\
0 & 0  &  0  &  0
\end{array}
\right)\, , \quad 
\varepsilon^{\alpha\beta}_{2}=\frac{1}{\sqrt{2}}\left( \begin{array}{cccc}
0 & 0  &  0  &  0   \\
0 & 0  &  1  &  0   \\
0 & 1  &  0  &  0  \\
0 & 0  &  0  &  0
\end{array}
\right)\, ;
\label{eq:83}
\end{equation}
so that the two complex combinations
\begin{equation}
\varepsilon^{\alpha\beta}_{\pm}(\bol{k})=\frac{1}{\sqrt{2}}\varepsilon^{\alpha\beta}_{1} 
\pm \frac{i}{\sqrt{2}}\varepsilon^{\alpha\beta}_{2}
\label{eq:84}
\end{equation}
represent states which have helicity $\pm 2$. Under rotation around the third axis
\begin{equation}
R(\va)=\left( \begin{array}{cccc}
1 & 0  &  0  &  0   \\
0 & \cos(\va)  &  \sin(\va)  &  0   \\
0 & -\sin(\va)  &  \cos(\va)  &  0  \\
0 & 0  &  0  &  1
\end{array}
\right)\, ,
\label{eq:85}
\end{equation}
they transform as follows
\begin{equation}
\varepsilon^{'\alpha\beta}_{\pm}(\bol{k})=R(\va)^{\alpha}_{\ \delta}\,  R(\va)^{\beta}_{\ \gamma}\, 
\varepsilon^{\delta\gamma}_{\pm}(\bol{k})=\big(R(\va)\, \varepsilon_{\pm}\, R^{T}(\va)\big)^{\alpha\beta} \, .
\label{eq:86}
\end{equation}
After multiplying out the matrices in~(\ref{eq:86}), we find
\begin{equation}
\varepsilon^{'}_{\pm}=e^{\pm 2i\va}\, \varepsilon_{\pm}
\label{eq:87}
\end{equation}
which means that the states have helicity $\pm2$ and represent free, physical gravitons.
The linear combinations~(\ref{eq:84}) imply that the operators  defined by
\begin{gather}
a_{\pm}(\bol{p})^{\dagger} := \frac{1}{\sqrt{2}} J_{-}(\bol{p})^{\dagger}  \mp i\, A^{12}(\bol{p})^{\dagger}\, ,\nonumber \\
\big[ a_{\pm}(\bol{k}), a_{\pm}(\bol{p})^{\dagger}\big]=\delta^{\sss (3)}(\bol{k}-\bol{p})\, , \quad 
\big[ a_{\pm}(\bol{k}), a_{\mp}(\bol{p})^{\dagger}\big]=0 \, ,
\label{eq:88}
\end{gather}
create physical $\pm2$-helicity  states from $|\Omega\rangle$:
\begin{equation}
|\Phi\rangle=\int\!\! d^{3}k\, \Big\{ \hat{f}_{+}(\bol{k}) a_{+}^{\dagger}(\bol{k}) +
                           \hat{f}_{-}(\bol{k}) a_{-}^{\dagger}(\bol{k})  \Big\} \, |\Omega\rangle \in \big( 
\mathcal{F}_{phys}\cap \mathcal{H}^{\sss (1)}\big)\, .
\end{equation}
These new operators, defined in Eq.~(\ref{eq:88}), will find an application in the next section.

\subsection{Hamilton Operators}

We compute the generators of the time evolutions for the fields $H^{\alpha\beta}(x)$ and $\Phi(x)$, respectively.
These free quantum fields satisfy the Heisenberg equations of motion
\begin{equation}
-i\dot H^{\alpha\beta}(x)=\big[\mathbf{H}_{\sss H},H^{\alpha\beta}(x)\big]\, , \quad 
-i\dot \Phi (x)=\big[ \mathbf{H}_{\sss \Phi},\Phi(x)\big]\, .
\label{eq:89}
\end{equation}
The  Hamilton operators are easily found by~(\ref{eq:50}) and read:
\begin{gather}
\mathbf{H}_{\sss H}=\int\! \!d^{3}p\, \omega( \bol{p}) \bigg[+ \sum_{\nu =0}^{3} 
{A^{\nu\nu}(\bol{p})}^{\dagger}A^{\nu\nu}(\bol{p}) 
+2 \sum_{i=1}^{3}{ A^{0i}(\bol{p})}^{\dagger}A^{0i}(\bol{p}) \nonumber \\
+2 \sum_{\substack{i,j=1\\i<j}}^{3}{ A^{ij}(\bol{p})}^{\dagger} A^{ij}(\bol{p}) \bigg]\, ,\nonumber \\
\mathbf{H}_{\sss \Phi}=
\frac{1}{4}\int\! \!d^{3}p\, \omega( \bol{p})\, a(\bol{p})^{\dagger} a(\bol{p})\, .
\label{eq:90}
\end{gather}
If we restrict these operators to the physical subspace ${\mathcal{F}_{phys}}$ and, in addition, 
choose the special reference frame as in Sec.~\ref{sec:physsub}, we find that the integrand of $\big( \mathbf{H}_{\sss H}  
+\mathbf{H}_{\sss \Phi} \big)$ reads
\begin{equation}
\omega\, \Big[ J_{-}^{\dagger}J_{-} +2 {A^{12}}^{\dagger} A^{12} \Big]\, .
\label{eq:91}
\end{equation}
This expression can be recast with~(\ref{eq:88}) into
\begin{equation}
\omega\, \Big[ a_{+}^{\dagger}a_{+} +a_{-}^{\dagger}a_{-} \Big] \, ,
\label{eq:92}
\end{equation}
which confirms that graviton states in ${\mathcal{F}_{phys}}$ have only two independent components,
the other eight  being unphysical.

\subsection{One-Graviton States in $\mathcal{F}_{phys}$}

As in Sec.~\ref{sec:fock}, we identify  one-graviton states $|\Phi\rangle$  with Lorentz
tensor potentials $\va^{\alpha\beta}(x)$ through Eq.~(\ref{eq:60}). 
For a state $|\Phi\rangle\in\mathcal{F}_{phys}$ the constraints
\begin{equation}
\va^{00}=0\, , \quad \sum_{i=1}^{3}\va^{ii}=0\, ,\quad \va^{0i}=0\, ,\quad \va^{ij}_{,j}=0
\label{eq:93}
\end{equation}
are satisfied. These eight conditions imply that $\va^{\alpha\beta}$ has the same form as the classical
polarization tensor~(\ref{eq:84}) with only two degrees of freedom.

\section{Krein Structure and Pseudo-Unitarity}\label{sec:krein}
\setcounter{equation}{0}

The drawback of the field representations~(\ref{eq:47}) and~(\ref{eq:48}) is that they do not have a
unitary implementation of Poincar\'e invariance. In order to remedy this defect, we introduce a new 
conjugation on the Fock space with respect to which the theory is unitary. 

\subsection{Poincar\'e Invariance and Krein Structure in the Graviton Sector}\label{sec:krein1}

Here we follow  the ideas and methods used in~\cite{scha} for the photon field case, by extending them 
to the more involved case of a  rank-2 tensor field. Considering the fields smeared out with real Schwartz test functions
\begin{equation}
H(f)=\int\!\!d^{4}x\, f_{\mu\nu}(x)H^{\mu\nu}(x)\, ,\quad \Phi(g)=\int\!\! d^{4}x \, g(x)\Phi(x)\, ;
\label{eq:95}
\end{equation}
then the Poincar\'e transformations of the test functions
\begin{equation}
f^{'}_{\mu\nu}(x^{'})=\Lambda_{\mu}^{\ \rho}\Lambda_{\nu}^{\ \sigma}\, f_{\rho\sigma}(\Lambda^{-1}(x-a))\, ,\quad
g^{'}(x^{'})=g(\Lambda^{-1}(x-a))\, ,
\label{eq:96}
\end{equation}
are lifted into Fock space by the definitions
\begin{gather}
\mathbf{U}(a,\Lambda)\, H(f)\, \mathbf{U}(a,\Lambda)^{-1}=H(f^{'})=\int\!\! d^{4}x\, f_{\mu\nu}(x)\big(\Lambda^{-1}\big)^{\mu}_{\ \rho}
\big(\Lambda^{-1}\big)^{\nu}_{\ \sigma}\, H^{\rho\sigma}(\Lambda x +a)\, ,\nonumber \\
\mathbf{U}(a,\Lambda)\, \Phi(g)\, \mathbf{U}(a,\Lambda)^{-1}=\Phi(g^{'})=\int\!\! d^{4}x\,  g(x)\, \Phi(\Lambda x +a)\, .
\label{eq:97}
\end{gather}
These lead to the transformation laws for the field operators:
\begin{gather}
\mathbf{U}(a,\Lambda)\,   H^{\mu\nu}(x)  \mathbf{U}(a,\Lambda)^{-1}=\big(\Lambda^{-1}\big)^{\mu}_{\ \rho}
\big(\Lambda^{-1}\big)^{\nu}_{\ \sigma}\, H^{\rho\sigma}(\Lambda x +a)\, , \nonumber \\
\mathbf{U}(a,\Lambda)\,  \Phi(x) \, \mathbf{U}(a,\Lambda)^{-1}=\Phi(\Lambda x+a)\, .
\label{eq:98}
\end{gather}
The problem is that the above representation $\mathbf{U}(a,\Lambda)$ is not unitary, because of the non-self-adjointness of
$H^{0i}(x)$ and of $\Phi(x)$.

Nevertheless, we can introduce another conjugation, `$\mathrm{K}$', in Fock space such that $\mathbf{U}(a,\Lambda)$ is
pseudo-unitary, \emph{i.e.} unitary with respect to the $\mathrm{K}$-conjugation:
\begin{equation}
\mathbf{U}(a,\Lambda)^{\mathrm{K}}= \mathbf{U}(a,\Lambda)^{-1} \, .
\label{eq:99}
\end{equation}
The new conjugation $\mathrm{K}$, which  defines the so-called Krein structure on $\mathcal{F}$, acts on the field
operators as 
\begin{equation}
\big( H^{\alpha\beta}(x) \big)^{\mathrm{K}}=\eta_{\sss H}\,{H^{\alpha\beta}(x)}^{\dagger}\, \eta_{\sss H}\, ,\quad
\big(\Phi(x)\big)^{\mathrm{K}}=\eta_{\sss \Phi}\, \Phi(x)^{\dagger}\, \eta_{\sss  \Phi} \, ,
\label{eq:100}
\end{equation}
where $\eta_{\sss H}$ and $\eta_{\sss  \Phi}$ are the Krein operators for the $H$-sector and $\Phi$-sector, respectively:
\begin{equation}
\eta_{\sss H}=\bigotimes_{i=1}^{3} (-1)^{\mathbf{N}_{0i}}\, , \quad \eta_{\sss \Phi}=(-1)^{\mathbf{N}_{\Phi}}\, ;
\label{eq:101}
\end{equation}
with the particle number operators for the $0i$-modes of the `$H$-graviton' and for the `scalar graviton' $\Phi$
\begin{equation}
\mathbf{N}_{0i}=2 \int\!\! d^{3}k\, A^{0i}(\bol{k})^{\dagger} A^{0i}(\bol{k})\, , \quad
\mathbf{N}_{\Phi}=\frac{1}{4}\int\!\! d^{3}k\, a(\bol{k})^{\dagger}a(\bol{k})\,  ;
\label{eq:102}
\end{equation}
which turned out to be unphysical, see  Sec.~\ref{sec:physsub}. This $\mathrm{K}$-conjugation shares all 
the properties of the $\dagger$-adjoint and the fields $H^{\alpha\beta}(x)$ and $\Phi(x)$ become self-conjugate 
with respect to the $\mathrm{K}$-conjugation:
\begin{equation}
\big( H^{\alpha\beta}(x) \big)^{\mathrm{K}}=H^{\alpha\beta}(x)\, , \quad \big(\Phi(x)\big)^{\mathrm{K}}=\Phi(x)\, ;
\label{eq:103}
\end{equation}
because the skew-adjointness with respect to the $\dagger$-adjoint is compensated by anti-commuting with 
$\eta_{\sss H}$ and $\eta_{\sss \Phi}$ respectively: Eq.~(\ref{eq:100}) for the absorption operators implies
\begin{gather}
a(\bol{p})^{\mathrm{K}}     =\eta_{\sss \Phi}\, a(\bol{p})^{\dagger}\,      \eta_{\sss \Phi}= -a(\bol{p})^{\dagger} \, , \nonumber \\
A^{0i}(\bol{p})^{\mathrm{K}}=\eta_{\sss H}   \, {A^{0i}(\bol{p})}^{\dagger}\, \eta_{\sss  H}  = -A^{0i}(\bol{p})^{\dagger} \, ,
\label{eq:104}
\end{gather}
from which the $\mathrm{K}$-self-conjugacy of the fields is evident:
\begin{gather}
H^{\alpha\beta}(x)=(2\pi)^{-3/2}\int\!\! \frac{d^{3}k}{\sqrt{2\omega (\bol{k})}} \Big( A^{\alpha\beta}(\bol{k})e^{-i\,k \cdot x}
            +A^{\alpha\beta}(\bol{k})^{\mathrm{K}} e^{+i\,k\cdot x}\Big)\, ,\nonumber \\
\Phi(x)=(2\pi)^{-3/2}\int\!\! \frac{d^{3}k}{\sqrt{2\omega (\bol{k})}}\Big( a(\bol{k})e^{-i\,k\cdot x}+
                      a(\bol{k})^{\mathrm{K}} e^{+i\,k\cdot x}\Big)\, .
\label{eq:105}
\end{gather}
From the physical point of view, the $\mathrm{K}$-conjugation is `indistinguishable' from the $\dagger$-adjoint, because
matrix elements between physical states are the same. For $|\Phi\rangle , |\Psi\rangle\in\mathcal{F}_{phys}$ we have
\begin{equation}
\langle\Phi|\big( H^{\alpha\beta}\big)^{\mathrm{K}}|\Psi\rangle=
\langle\Phi| \eta_{\sss H}\,{H^{\alpha\beta}}^{\dagger}\, \eta_{\sss H}   |\Psi\rangle=
\langle \eta_{\sss H}\,\Phi|{H^{\alpha\beta}}^{\dagger}|\eta_{\sss H}\, \Psi\rangle
=\langle\Phi|{H^{\alpha\beta}}^{\dagger}|\Psi\rangle\ ,
\label{eq:106}
\end{equation}
because the number of unphysical $0i$-gravitons, $i=1,2,3$, is zero for states in $\mathcal{F}_{phys}$. The  same holds for
$\Phi$. Therefore the lack of self-adjointness in  $H^{\alpha\beta}(x)$ and in  $\Phi(x)$ does not cause any problem.

\subsection{Krein Structure in the Ghost Sector}\label{sec:krein2}

We extend the $\dagger$-adjoint to the K-conjugation also in the ghost sector.
The fully investigation of the fermionic ghost vector fields is not carried out here,
we note only that micro-causality, Eq.~(\ref{eq:29}), is not broken, because of the unusual mixture
of creation and absorption operators in the ghost free field representations~(\ref{eq:72}).
The ghost fields have non-causal anti-commutation relations with their adjoints  so that they
do not fulfil the assumptions of the spin-statistic theorem and therefore they are allowed to escape its
consequences~\cite{kra3}.

For later use (sec.~\ref{sec:unit}), we now require the ghost field $u^{\alpha}(x)$ to be 
K-self-conjugate and the anti-ghost field $\tilde{u}^{\alpha}(x)$ to be K-skew-conjugate:
\begin{equation}
\big( u^{\alpha}(x)\big)^{\mathrm{K}} = u^{\alpha}(x)\, , \quad    
\big( \tilde{u}^{\alpha}(x)\big)^{\mathrm{K}}=-  \tilde{u}^{\alpha}(x)\, .
\label{eq:107}
\end{equation}
These requirements imply that there must exist a Krein operator $\eta_{\sss G}$ acting on the ghost Fock space 
so that we have for the creation and absorption operators
\begin{equation}
\begin{split}
\big( b^{\mu}(\bol{p})\big)^{\mathrm{K}}&=\eta_{\sss G}\, b^{\mu}(\bol{p})^{\dagger}\,  \eta_{\sss G}=\begin{cases}
\big( b^{0}(\bol{p})\big)^{\mathrm{K}}=\eta_{\sss G}\, b^{0}(\bol{p})^{\dagger}\,  \eta_{\sss G}=-c^{0}(\bol{p})^{\dagger}\, , \\
\big( b^{i}(\bol{p})\big)^{\mathrm{K}}=\eta_{\sss G}\, b^{i}(\bol{p})^{\dagger}\,  \eta_{\sss G}=+c^{i}(\bol{p})^{\dagger}\, ,
\end{cases}  \\
\big( c^{\mu}(\bol{p})\big)^{\mathrm{K}}&=\eta_{\sss G}\, c^{\mu}(\bol{p})^{\dagger}\,  \eta_{\sss G}=\begin{cases}
\big( c^{0}(\bol{p})\big)^{\mathrm{K}}=\eta_{\sss G}\, c^{0}(\bol{p})^{\dagger}\,  \eta_{\sss G}=-b^{0}(\bol{p})^{\dagger}\, , \\
\big( c^{i}(\bol{p})\big)^{\mathrm{K}}=\eta_{\sss G}\, c^{i}(\bol{p})^{\dagger}\,  \eta_{\sss G}=+b^{i}(\bol{p})^{\dagger}\, .
\end{cases}
\label{eq:108}
\end{split}
\end{equation}
Again, it is the unusual mixture of $b^{\mu}$ and $c^{\nu}$ operators in~(\ref{eq:72}) that makes~(\ref{eq:107}) and~(\ref{eq:108})
possible.

Extending the $\dagger$-adjoint to the K-conjugation we obtain also for the ghost fields the more symmetric form
\begin{equation}
\begin{split}
u^{\nu}(x)&=(2\pi)^{-3/2}\int\!\!\frac{d^{3}p}{\sqrt{2\omega(\bol{p})}} \Big( +b^{\nu}(\bol{p})e^{-i\,p \cdot x}+
 {b^{\nu}(\bol{p})}^{\mathrm{K}} e^{i\,p\cdot x} \Big)\,  , \\
\tilde{u}^{\nu}(x)&=(2\pi)^{-3/2}\int\!\!\frac{d^{3}p}{\sqrt{2\omega(\bol{p})}} \Big( -c^{\nu}(\bol{p})e^{-i\,p\cdot x}
+{c^{\nu}(\bol{p})}^{\mathrm{K}} e^{i\,p\cdot x} \Big)\,  .
\label{eq:109}
\end{split}
\end{equation}
The construction of $\eta_{\sss G}$, which performs the transformations~(\ref{eq:107}) requires more work. Let us define 
the following conserved currents
\begin{align}
j^{\mu}_{\sss N}(x)&:= i\, :\!u^{\alpha}(x)^{\dagger}{\stackrel{\longleftrightarrow}{ \partial_{x}^{\mu} }}u^{\beta}(x)\! :
                               \eta_{\alpha\beta}\, ,  & 
j^{\mu}_{\sss G}(x)&:= i\, :\!\tilde{u}^{\alpha}(x){\stackrel{\longleftrightarrow}{ \partial_{x}^{\mu} }}u^{\beta}(x)\! :
                      \eta_{\alpha\beta}\, ,\nonumber \\
j^{\mu}_{u}(x)&:= i\, :\!u^{\alpha}(x){\stackrel{\longleftrightarrow}{ \partial_{x}^{\mu} }}u^{\beta}(x)\!:
                       \eta_{\alpha\beta}\, ,&
j^{\mu}_{\tilde{u}}(x)&:= i\, :\!\tilde{u}^{\alpha}(x){\stackrel{\longleftrightarrow}{ \partial_{x}^{\mu} }}
                      \tilde{u}^{\beta}(x)\!: \eta_{\alpha\beta}\, .
\label{eq:110}
\end{align}
From these currents we compute the conserved ghost charges by inserting the free field representations~(\ref{eq:72}) and
using the anti-commutation relations~(\ref{eq:71}):
\begin{equation}
N_{\sss G}:=\int\limits_{x^{0}=const}\!\! d^{3}x\, j^{0}_{\sss N}(x)=N_{\sss G}^{\sss (0)}-\sum_{i=1}^{3}N_{\sss G}^{\sss (i)}\ ,
                \quad N_{\sss G}^{\sss (\mu)}:=
    \int \!\! d^{3}p\, \big({b^{\mu}}^{\dagger}b^{\mu}+{c^{\mu}}^{\dagger}c^{\mu}\big)\, , 
\end{equation}
\begin{equation}
Q_{\sss G}:=\int\limits_{x^{0}=const}\!\! d^{3}x\, j^{0}_{\sss G}(x)=-\sum_{\mu=0}^{3} Q_{\sss G}^{\sss (\mu)}\, ,
\quad Q_{\sss G}^{\sss (\mu)}:=
     \int \!\!d^{3}p\, \big({b^{\mu}}^{\dagger}b^{\mu}-{c^{\mu}}^{\dagger}c^{\mu}\big)\, , 
\end{equation}
\begin{gather}
\Gamma_{\sss G}:=\frac{1}{2}\int\limits_{x^{0}=const}\!\! d^{3}x\,
    \big( j^{0}_{u}(x)-j^{0}_{\tilde{u}}(x)\big) =-\sum_{\mu=0}^{3}\Gamma_{\sss G}^{\sss (\mu)}\, ,\nonumber \\
       \Gamma_{\sss G}^{\sss (\mu)}:=
    \int \!\! d^{3}p\, \big({b^{\mu}}^{\dagger}c^{\mu}+{c^{\mu}}^{\dagger}b^{\mu}\big)\, , 
\end{gather}
\begin{gather}
\Omega_{\sss G}:=\frac{i}{2}\int\limits_{x^{0}=const}\!\! d^{3}x\,
    \big( j^{0}_{u}(x)+j^{0}_{\tilde{u}}(x)\big) =-\sum_{\mu=0}^{3}\Omega_{\sss G}^{\sss (\mu)}\, ,\nonumber \\
 \Omega_{\sss G}^{\sss (\mu)}:=
    \frac{1}{i}\int \!\! d^{3}p\, \big({b^{\mu}}^{\dagger}c^{\mu}-{c^{\mu}}^{\dagger}b^{\mu}\big)\, ;
\label{eq:111}
\end{gather}
where $Q_{\sss G}^{\sss (\mu)}$ is the $\mu$-ghost charge, $\Gamma_{\sss G}^{\sss (\mu)}$ and $\Omega_{\sss G}^{\sss (\mu)}$ are two
transfer operators which replace one $\mu$-anti-ghost by a $\mu$-ghost component and vice versa with a different 
relative sign; the $\mu$-ghost number $N_{\sss G}^{\sss (\mu)}$
can be used to form the Hamilton operator
\begin{equation}
\mathbf{H}_{\sss G}=\sum_{\mu =0}^{3} N_{\sss G}^{\sss (\mu)}\, ,
\label{eq:112}
\end{equation}
which implements the Heisenberg time evolutions of the ghost and anti-ghost fields
\begin{equation}
-i\, \dot u^{\mu}(x)=\big[ \mathbf{H}_{\sss G} , u^{\mu}(x) \big]\, ,\quad 
-i\, \dot{\tilde{u}}^{\mu}(x)=\big[\mathbf{H}_{\sss G} , \tilde{u}^{\mu}(x)\big]\, .
\label{eq:113}
\end{equation}
The charges in~(\ref{eq:111}) and the gauge charge $Q$ satisfy
\begin{equation}
Q^{\mathrm{K}}=Q\, ,\quad Q_{\sss G}^{\mathrm{K}}=-Q_{\sss G}\, ,\quad N_{\sss G}^{\mathrm{K}}=N_{\sss G}\, ,\quad
\Gamma_{\sss G}^{\mathrm{K}}=\Gamma_{\sss G}\, ,\quad \Omega_{\sss G}^{\mathrm{K}}=-\Omega_{\sss G}\, .
\label{eq:114}
\end{equation}
Finally, the Krein operator $\eta_{\sss G}$ that implements~(\ref{eq:108}) is given by
\begin{equation}
\eta_{\sss G}=\exp\Big( i\,\frac{\pi}{2}\big( N_{\sss G} -\Gamma_{\sss G}\big)\Big)\, .
\label{eq:115}
\end{equation}
The motivation for this result can be traced back to the algebraic analysis of ghost fields within  non-Abelian gauge 
theories in~\cite{kra3}. The proof that the Krein operator given in~(\ref{eq:115}) produces the desired 
relations~(\ref{eq:108}) is just a simple consistency check by means of the Baker-Hausdorff formula and of the
commutators of  the ghost charges $N_{\sss G}$ and $\Gamma_{\sss G}$ with ${b^{\mu}}^{\dagger}$ and 
${c^{\mu}}^{\dagger}$.

\section{Gauge Transformation and Poincar\'e Invariance}\label{sec:poinc}

\setcounter{equation}{0}

The representation of Poincar\'e transformations for Fock space operators by means of the pseudo-unitary
$\mathbf{U}(a,\Lambda)$ cannot be the correct implementation of physical relativistic invariance, that is
of the fact that physical predictions have to be independent from the choice of reference frames which are moving
uniformly to each other.

A Poincar\'e transformation must give rise to a unitary mapping between physical states in $\mathcal{F}_{phys}$.
This unitary mapping is realized if we take advantage of the  gauge freedom still present in the theory.

We analyze this issue for the case of one-graviton states, the generalization to many-graviton states is
straightforward although very cumbersome. First we need some technical preparation.

\subsection{One-Graviton States and Hilbert-Einstein Subspace 
$\mathcal{F}_{\sss{HE}}$}

Although the Fock space for gravitons is constructed with the help of the $B^{\alpha\beta}$-operators, 
Sec.~\ref{sec:fock}, for the purpose of this section it is convenient to use the $A^{\alpha\beta}$-operators and
the $a$-operator of Eq.~(\ref{eq:50}) in order to describe one-graviton states. 

Since the $A^{\alpha\beta}$'s are not independent, because of $A^{\alpha}_{\;\alpha}=0$, another constraint on 
the Lorentz tensor potential $\va^{\alpha\beta}$ has to be imposed.

We consider one-graviton states (skipping the time dependence) of the form
\begin{equation}
\begin{split}
|\Phi\rangle&=\big( A^{\dagger}(\va) +a^{\dagger}(f)\big)\, |\Omega\rangle \\
            &=\int\!\!d^{3}k\,\Big( \sum_{\mu,\nu=0}^{3} \hat{\va}^{\mu\nu}(\bol{k})\, A^{\mu\nu}(\bol{k})^{\dagger}
+\frac{1}{4}\hat{f}(\bol{k})\, a(\bol{k})^{\dagger} \Big)\, |\Omega\rangle\, ,
\label{eq:116}
\end{split}
\end{equation}
and represent it through the identification between a Fock space vector, on the one hand, and a traceless symmetric
tensor potential together with a scalar potential, on the other hand:
\begin{equation}
|\Phi\rangle\in\mathcal{H}^{\sss (1)} \longleftrightarrow  \begin{cases} & \va^{\mu\nu}(x)\;\text{\ with } 
\eta_{\rho\sigma}\va^{\rho\sigma}(x)=0\ ,
\, \Box \va^{\mu\nu}(x)=0 \\
&\text{and } f(x)\text{ with \ } \Box f(x)=0\, .\end{cases}
\end{equation}
The scalar product in the scalar sector is 
\begin{equation}
(f,g)_{scal}:=\int\!\!d^{3}p\,\hat{f}(\bol{p})^{\ast} \hat{g}(\bol{p})\, ,
\label{eq:117}
\end{equation}
and in the tensor sector reads
\begin{equation}
(\psi,\va)_{\sss E}:=\sum_{\mu,\nu=0}^{3} i\int \!\!d^{3}x\,\Big[\psi^{\mu\nu}(x)^{\ast}\partial_{t}\va^{\mu\nu}(x)-
\big(\partial_{t}\psi^{\mu\nu}(x)\big)^{\ast}\va^{\mu\nu}(x)\Big]\, .
\label{eq:117.1}
\end{equation}
With the Fourier representation of the tensor potential
\begin{equation}
\va^{\mu\nu}(t,\bol{x})=(2\pi)^{-3/2}\int\!\! \frac{d^{3}p}{\sqrt{2\omega(\bol{p})}}\,\hat{\va}^{\mu\nu}(\bol{p})\,
e^{-i(\omega(\bol{p})\, t-\bol{p}\cdot\bol{x})} \,  ,
\label{eq:117.2}
\end{equation}
the scalar product becomes
\begin{equation}
(\psi,\va)_{\sss E}=\sum_{\mu,\nu=0}^{3} \int \!\!d^{3}p\, \hat{\psi}^{\mu\nu}(\bol{p})^{\ast}
\hat{\va}^{\mu\nu}(\bol{p})\, ;
\label{eq:118}
\end{equation}
this formula already appeared in Eq.~(\ref{eq:61}). The sum over $\mu$ and $\nu$ in~(\ref{eq:117.1}) is not a Minkowski
scalar product, but it can be written in a Lorentz covariant way as a surface integral
\begin{equation}
(\psi,\va)_{\sss E}=\sum_{\mu,\nu=0}^{3} i\int\limits_{x^{0}=const} \!\!d\sigma^{\alpha}(x)\,
\Big[\psi^{\mu\nu}(x)^{\ast}\partial_{\alpha}^{x}\va^{\mu\nu}(x)-
\big(\partial_{\alpha}^{x}\psi^{\mu\nu}(x)\big)^{\ast}\va^{\mu\nu}(x)\Big]\, ,
\label{eq:119}
\end{equation}
that can be taken over an arbitrary smooth space-like surface as a consequence of Gauss' theorem and of the wave equation
for the tensor potentials.

The Fock space scalar product between one-graviton states $|\Psi\rangle$ and $|\Phi\rangle$, characterized
by $\{ \psi^{\mu\nu}(x),f(x)\}$ and $\{ \va^{\mu\nu}(x),g(x)\}$, respectively, can be computed by means of 
Eq.~(\ref{eq:50}):
\begin{equation}
\langle\Psi|\Phi\rangle=\int\!\! d^{3}k \int\!\!d^{3}p\bigg(\sum_{\alpha,\beta,\atop \mu,\nu=0}^{3} 
\hat{\psi}^{\alpha\beta}(\bol{k})^{\ast}\, \tilde{t}^{\alpha\beta\mu\nu}\,\hat{\va}^{\mu\nu}(\bol{p})+
\frac{1}{4}\hat{f}(\bol{k})^{\ast} \hat{g}(\bol{p})\bigg)\,\delta^{\sss (3)}(\bol{k}-\bol{p}) \, .
\label{eq:120}
\end{equation}
Because of the trace condition $\psi^{\sigma}_{\ \sigma}=\va^{\sigma}_{\ \sigma}=0$ we obtain
\begin{equation}
\begin{split}
\langle\Psi|\Phi\rangle&=\int\!\! d^{3}k \,\bigg(\sum_{\mu,\nu=0}^{3} 
\hat{\psi}^{\mu\nu}(\bol{k})^{\ast} \hat{\va}^{\mu\nu}(\bol{k})+
\frac{1}{4}\hat{f}(\bol{k})^{\ast} \hat{g}(\bol{k})\bigg)\\
&=(\psi,\phi)_{\sss E} +\frac{1}{4}(f,g)_{scal}\, .
\label{eq:121}
\end{split}
\end{equation}
A one-graviton state $|\Psi\rangle\in\mathcal{F}_{phys}$ certainly satisfies
\begin{equation}
A^{0\mu}(\bol{p})|\Psi\rangle=A_{\parallel}^{\mu}(\bol{p})|\Psi\rangle=A^{\nu}(\bol{p})_{\nu}|\Psi\rangle
=a(\bol{p})|\Psi\rangle=0\, ,
\end{equation}
because of Eq.~(\ref{eq:79}). Therefore the potentials $\{\psi^{\mu\nu}(x),f(x)\}$  are constrained as follows
\begin{equation}
\psi^{00}=\psi^{0i}=\sum_{i=1}^{3} \psi^{ii}=\psi^{ij}_{,j}=f=0\, .
\label{eq:122}
\end{equation}
These nine  conditions reduce the number of independent components of $\psi^{\mu\nu}(x)$ and $f(x)$ from eleven
to two.

Although~(\ref{eq:122}) depends on the reference frame, $\mathcal{F}_{phys}$ is independent of such a choice, 
because one can show that the identification $\mathcal{F}_{phys}=\ker Q / \mathrm{ran}\;Q$
is equivalent to Eq.~(\ref{eq:70}) and independent of a reference frame~\cite{ym4}.

In addition to the physical subspace $\mathcal{F}_{phys}$, we introduce the Hilbert--Einstein subspace
$\mathcal{F}_{\sss HE}$ of states which satisfy the Hilbert-Einstein gauge condition
\begin{equation}
\big( H^{\alpha\beta}(x)_{,\beta}+\frac{1}{4}\Phi(x)^{,\alpha}\big)^{\sss (-)}\,|\Psi\rangle=0\, ,
\label{eq:123}
\end{equation}
for a one-graviton state $|\Psi\rangle$ with the representation~(\ref{eq:116}), this means
\begin{equation}
\va^{\alpha\beta}(x)_{,\beta}+\frac{1}{4}f(x)^{,\alpha} =0\, .
\label{eq:124}
\end{equation}
Obviously, states in  $\mathcal{F}_{phys}$ are automatically in $\mathcal{F}_{\sss HE}$.

\subsection{Poincar\'e Invariance and Gauge Projection}\label{sec:proj}

The aim of this section is  to show that for $|\Psi\rangle$ and $|\Phi\rangle$ in $\mathcal{F}_{phys}$,
the scalar product $\langle\Psi|\Phi\rangle$ remains invariant under a Poincar\'e transformation
$x\to x^{'}=\Lambda x +a$ in $\mathcal{P}_{+}^{\uparrow}$, provided that the tensor potentials undergo a 
gauge transformation along a fibre of gauge equivalent states, whereas the scalar potentials remain 
unaffected.

For physical states, $\langle\Psi|\Phi\rangle$ can be written as a Minkowski scalar product
\begin{equation}
\langle\Psi|\Phi\rangle=\sum_{\mu,\nu=0}^{3} \int\!\! d^{3}k  \,
\hat{\psi}^{\mu\nu}(\bol{k})^{\ast}\, \hat{\va}^{\mu\nu}(\bol{k})=\int\!\! d^{3}k \, 
\hat{\psi}^{\mu\nu}(\bol{k})^{\ast}\, \hat{\va}_{\mu\nu}(\bol{k})=:(\psi,\va)_{\sss M}
\label{eq:125}
\end{equation}
because $\psi^{\mu\nu}$ and $\va^{\mu\nu}$ have vanishing $0i$-components.

Under a Poincar\'e transformation~(\ref{eq:96}), the states $|\Psi\rangle$ and $|\Phi\rangle$ are `rotated out'
of $\mathcal{F}_{phys}$, this means that the tensor potentials acquire non-vanishing $0\mu$-components, whereas
\begin{align}
\psi^{\sigma}_{\ \sigma}&=0\longrightarrow\psi^{'\sigma}_{\ \sigma}=0\, , &
\va^{\lambda}_{\ \lambda}&=0\longrightarrow\va^{'\lambda}_{\ \lambda}=0\, , \\    
f&=0\longrightarrow f^{'}=0\, , & g&=0\longrightarrow g^{'}=0\, .
\label{eq:126}
\end{align}
Nevertheless, $(\psi,\va)_{\sss M}$ remains invariant, because the Minkowski product now  renders the expression in
Eq.~(\ref{eq:125}) invariant.

The states $|\Psi^{'}\rangle$ and $|\Phi^{'}\rangle$ are still in $\mathcal{F}_{\sss HE}$, 
because the potentials still satisfy
\begin{equation}
\psi^{'\alpha\beta}_{\ ,\beta}+\frac{1}{4}f^{' ,\alpha} =0\, ,\quad
\va^{'\alpha\beta}_{\ ,\beta}+\frac{1}{4}g^{',\alpha} =0\, ,
\label{eq:127}
\end{equation}
as a consequence of the Poincar\'e invariance of the Hilbert--Einstein gauge condition~(\ref{eq:124}) and,
because of Eq.~(\ref{eq:126}), we arrive at $\psi^{'\alpha\beta}_{\ ,\beta}=0$ and $\va^{'\alpha\beta}_{\ ,\beta}=0$.

Now, we would like to transform the expression for the Poincar\'e transformed Fock space scalar product 
$\langle\Psi^{'}|\Phi^{'}\rangle$, namely $(\psi^{'},\va^{'})_{\sss M}$, back into a  Euclidean sum, which defines
our scalar product in the tensor sector, but this is not possible being now 
$\psi^{' 0i}$ and $\va^{' 0i}$ different from zero.

By means of a gauge transformation, it is possible to bring the states $|\Psi^{'}\rangle$ and $|\Phi^{'}\rangle$ 
back into $\mathcal{F}_{phys}$, this means that  the potentials undergo the gauge transformations:
 $\psi^{'\alpha\beta}\to\tilde{\psi}^{\alpha\beta}$
and $\va^{'\alpha\beta}\to\tilde{\va}^{\alpha\beta}$ with $\tilde{\psi}^{0\alpha}=\tilde{\va}^{0 \alpha}=0,\,\forall
\alpha=0,\ldots,3$. These gauge transformations are  within the Hilbert--Einstein class and  read
\begin{equation}
\begin{split}
&\begin{cases}
&\psi^{'\alpha\beta}\longrightarrow \psi^{'\alpha\beta}=\tilde{\psi}^{\alpha\beta}+
      \tilde{t}^{\alpha\beta\gamma\delta}\,u^{\gamma}_{\; ,\delta}\, , \\
&f^{'}\longrightarrow f^{'}=\tilde{f}-u^{\gamma}_{\;, \gamma}\, ,
\end{cases}\\
&\begin{cases}
&\va^{'\alpha\beta}\longrightarrow \va^{'\alpha\beta}=\tilde{\va}^{\alpha\beta}+
      \tilde{t}^{\alpha\beta\gamma\delta}\,v^{\gamma}_{\; ,\delta}\, , \\
&g^{'}\longrightarrow g^{'}=\tilde{g}-v^{\gamma}_{\;, \gamma}\, ,
\end{cases}
\label{eq:128}
\end{split}
\end{equation}
where the  vector fields $u^{\gamma}(x)$ and $v^{\delta}(x)$ satisfy the wave equation.
This means that each $|\Psi^{'}\rangle\in\mathcal{F}_{\sss HE}$ is connected with a unique
$|\tilde{\Psi}\rangle\in\mathcal{F}_{phys}$ by a fibre of gauge equivalent states. Since
$f^{'}=g^{'}=0$ (Poincar\'e transformations do not alter the scalar part) and $\tilde{f}=\tilde{g}=0$
($|\tilde{\Psi}\rangle$ and $|\tilde{\Phi}\rangle$ are in $\mathcal{F}_{phys}$), the transformations
in Eq.~(\ref{eq:128}) are automatically restricted to the case in which 
$ u^{\gamma}_{\;, \gamma}=v^{\gamma}_{\;, \gamma}=0$.
Therefore, using Eq.~(\ref{eq:128}), the Minkowski scalar product $(\psi^{'},\va^{'})_{\sss M}$ becomes
\begin{equation}
\begin{split}
(\psi^{'},\va^{'})_{\sss M}&=(\tilde{\psi},\tilde{\va})_{\sss M}+ (\tilde{\psi}^{\mu\nu},v^{\mu}_{\; ,\nu})_{\sss M} +
       (u^{\mu}_{\; ,\nu},\tilde{\va}^{\mu\nu})_{\sss M}\\
&\quad +\frac{1}{2}(u^{\nu}_{\; ,\mu},v^{\mu}_{\; ,\nu})_{\sss M}
        +\frac{1}{2}(u^{\mu}_{\; ,\nu},v^{\mu}_{\; ,\nu})_{\sss M}\, .
\label{eq:129}
\end{split}
\end{equation}
The second and third terms on the right side vanish after 3-dimensional partial integration because of 
$\tilde{\psi}^{\mu\nu}_{\ ,\nu}=\tilde{\va}^{\mu\nu}_{\ ,\nu}=0$. The fourth term vanishes because of
$ u^{\gamma}_{\;, \gamma}=v^{\gamma}_{\;, \gamma}=0$ and the last term also vanishes because it can be put
in the form
\begin{equation}
(\partial_{\nu} u^{\mu},\partial_{\nu} v^{\mu})_{\sss M}=\frac{1}{2}\,\partial_{0}^2 \, (u^{\mu},v^{\mu})_{\sss M}=0\ ,
\label{eq:130}
\end{equation}
and vanishes because the scalar product between two solutions of the wave equation is constant in time.

Since now $\tilde{\psi}^{0\alpha}=\tilde{\va}^{0 \alpha}=0,\,\forall \alpha=0,\ldots,3$, we arrive at
\begin{multline}
\langle\Psi|\Phi\rangle  \stackrel{(\ref{eq:118})}{=}  (\psi,\va)_{\sss E}\stackrel{(\ref{eq:125})}{=}(\psi,\va)_{\sss M} 
 \stackrel{\text{Poincar\'e}}{=}    (\psi^{'},\va^{'})_{\sss M} \\
 \stackrel{\mathrm{gauge\ tr.}}{=}    (\tilde{\psi},\tilde{\va})_{\sss M}+\mathrm{ vanishing\  terms}
\stackrel{(\ref{eq:122})}{=}    (\tilde{\psi},\tilde{\va})_{\sss E}\stackrel{(\ref{eq:118})}{=}
    \langle\tilde{\Psi}|\tilde{\Phi}\rangle\ ,
\label{eq:131}
\end{multline}
with $|\tilde{\Psi}\rangle$ and $|\tilde{\Phi}\rangle$ in $\mathcal{F}_{phys}$.
Eq.~(\ref{eq:131}) contains the desired unitary mapping in  $\mathcal{F}_{phys}$:
\begin{equation}
\langle\tilde{\Psi}|\tilde{\Phi}\rangle =\langle\tilde{ \mathbf{U}} \Psi |\tilde{ \mathbf{U}} \Phi \rangle=
\langle \Psi |\tilde{ \mathbf{U}}^{\dagger} \tilde{ \mathbf{U}}| \Phi \rangle=\langle{\Psi}|{\Phi}\rangle\, ,
\label{eq:132}
\end{equation}
with $\tilde{\mathbf{U}}^{\dagger}=\tilde{\mathbf{U}}^{-1}$. Therefore the pseudo-unitary Fock space implementation
of Poincar\'e transformations cooperates with a gauge projection to give a unitary transformation 
$\tilde{\mathbf{U}}$ in $\mathcal{F}_{phys}$.

\section{Unitarity}\label{sec:unit}
\setcounter{equation}{0}

Unitarity is the root of the probabilistic interpretation of quantum mechanics and of $S$-matrix theory.
In quantum gravity, as in non-Abelian gauge theories, unitarity is a very important property, which one 
has to prove, because the Fock space $\mathcal{F}$ contains a lot of unphysical states (ghost states  
and additional polarization states of the $H$-graviton and of the $\Phi$-graviton).

Nevertheless, there exist a physical subspace $\mathcal{F}_{phys}$, Eq.~(\ref{eq:70}),  with a positive 
definite scalar product, Eq.~(\ref{eq:121}), such that the $S$-matrix restricted to $\mathcal{F}_{phys}$
is unitary, whereas unitarity does not hold on the entire Fock space $\mathcal{F}$.

In this section we arrange  the proof of unitarity for  non-Abelian gauge theories~\cite{ym4},~\cite{asd}
to the quantum gravity case.

\subsection{{\boldmath{$S$}}-Matrix Unitarity on {${\mathcal{F}_{phys}}$}}

The total Krein operator $\eta=\eta_{\sss H}\otimes\eta_{\sss \Phi}\otimes\eta_{\sss G}$ defines the K-conjugation
acting as
\begin{equation}
\mathcal{O}^{\mathrm{K}}=\eta\, \mathcal{O}^{\dagger}\,\eta\, ,
\label{eq:3.0}
\end{equation}
on the Fock space  operator $\mathcal{O}$. Using~(\ref{eq:103}) and~(\ref{eq:107}), we find that the first 
order interaction $T_{1}^{h+u}(x)$, Eq.~(\ref{eq:26}) and Eq.~(\ref{eq:35}), is skew-conjugate with respect 
to the K-conjugation:
\begin{equation}
\big(T_{1}^{h+u}(x)\big)^{\mathrm{K}}=-T_{1}^{h+u}(x)=\tilde{T}_{1}^{h+u}(x)\, ,
\label{eq:3.1}
\end{equation}
where $\tilde{T}_{1}(x)$ is the first term in the expansion of the inverse $S$-matrix (see below), and is Poincar\'e invariant:
\begin{equation}
\mathbf{U}(a,\Lambda)\, T_{1}^{h+u}(x)\, \mathbf{U}(a,\Lambda)^{-1}=T_{1}^{h+u}(\Lambda x +a)\, ,
\label{eq:3.1.1}
\end{equation}
where $\mathbf{U}(a,\Lambda)$ is pseudo-unitary, Eq.~(\ref{eq:99}).

By induction, these properties hold for the $n$-point operator valued distributions $T_{n}$:
\begin{gather}
T_{n}(x_{1},\ldots,x_{n})^{\mathrm{K}}=\tilde{T}_{n}(x_{1},\ldots,x_{n}) \, , \nonumber \\
\mathbf{U}(a,\Lambda)\, T_{n}(x_{1},\ldots,x_{n})\, \mathbf{U}(a,\Lambda)^{-1}=
T_{n}(\Lambda x_{1}+a,\ldots,\Lambda x_{n}+a)\, ,
\label{eq:3.2}
\end{gather}
where $\tilde{T}_{n}$ is the $n$-point distribution belonging to $S(g)^{-1}$, the inverse $S$-matrix
\begin{equation}
S(g)^{-1}={\mathbf 1}+\sum_{n=1}^{\infty}\frac{1}{n!} \int\!\! d^{4}x_{1}
\ldots d^{4}x_{n}\, \tilde{T}_{n}(x_{1},\ldots,
x_{n})\,g(x_{1})\cdot\ldots\cdot g(x_{n})\, .
\label{eq:3.3}
\end{equation}
In the inductive construction of the $T_{n}$'s, these properties can get lost in the process of distribution splitting
only, Eq.~(\ref{eq:1.8}), but if the normalization constants $C_{a}$ are chosen in a suitable way, pseudo-unitarity
and Poincar\'e invariance go over from the first order~(\ref{eq:3.1}) and~(\ref{eq:3.1.1}) 
to higher orders~(\ref{eq:3.2}),~\cite{scha}.

The first equation in~(\ref{eq:3.2}) is the perturbative version of $S$-matrix pseudo-unitarity (with a real test 
function $g$):
\begin{equation}
S(g)^{\mathrm{K}}=S(g)^{-1}\, .
\label{eq:3.4}
\end{equation}
Unitarity of $S(g)$ is changed into pseudo-unitarity because of the unphysical degrees of freedom present in the theory
and because of the fact that the fields $H^{0i}$ and $\Phi$ are not Hermitian. Only after the elimination of these
two obstacles, by means of projecting onto $\mathcal{F}_{phys}$, we are able to show unitarity on the physical subspace,
by which we mean the heuristic  equation
\begin{equation}
\lim_{g\to 1}\Big[ P_{phys}\,S(g)^{\dagger}\,P_{phys}\Big]\,\Big[P_{phys}\, S(g)\, P_{phys}\Big]=P_{phys}\, ,
\label{eq:3.4.1}
\end{equation}
where $P_{phys}$ stands for the projection operator onto  $\mathcal{F}_{phys}$.
Perturbatively, Eq.~(\ref{eq:3.4.1}) goes over into
\begin{equation}
\tilde{T}_{n}^{P}(X)=P_{phys}\, T_{n}(X)^{\dagger}\, P_{phys} +\text{sum of divergences}\,,
\label{eq:3.5}
\end{equation}
where $X=\{x_{1},\ldots,x_{n}\}$ and $\tilde{T}_{n}^{P}(X)$ is the $n$-point distribution of the $S$-matrix
inverted on $\mathcal{F}_{phys}$:
\begin{multline}
\big( P_{phys}\, S(g)\, P_{phys}\big)^{-1}=P_{phys}+\sum_{n=1}^{\infty}\frac{1}{n!} \int\!\! d^{4}x_{1}
\ldots d^{4}x_{n}\, \tilde{T}_{n}^{P}(x_{1},\ldots,
x_{n})\cdot \\ \cdot g(x_{1})\cdot\ldots\cdot g(x_{n})\, .
\label{eq:3.6}
\end{multline}
The divergences appearing on the right side of~(\ref{eq:3.5}) do not contribute in the adiabatic limit
$g\to 1$. Our aim is to prove Eq.~(\ref{eq:3.5}). According to the theory of the inductive construction of 
the $S$-matrix~\cite{scha}, $\tilde{T}_{n}^{P}(X)$ is given by the following sum over subsets of $X$
\begin{equation}
\tilde{T}_{n}^{P}(X)=\sum_{r=1}^{n} (-1)^{r}\!\!\!\sum_{\text{perm. of r}\atop\text{ partition of X}} 
 P_{phys}\, T_{n_{1}}(X_{1})\,P_{phys}\ldots P_{phys}\,T_{n_{r}}(X_{r})\,P_{phys} \, .
\label{eq:3.7}
\end{equation}
Using perturbative gauge invariance to the $n$-th order, Eq.~(\ref{eq:37})
\begin{equation}
\big[ Q,T_{n}(X)\big]=\text{sum of divergences}\, ,
\end{equation}
we can get rid of all the internal physical projections (see Sec.~\ref{sec:lemma}) so that
\begin{equation}
\begin{split}
\tilde{T}_{n}^{P}(X)&=\sum_{r=1}^{n} (-1)^{r}\!\!\!
\sum_{\text{perm. of r}\atop\text{ partition of X}}\, P_{phys}
T_{n_{1}}(X_{1})\ldots T_{n_{r}}(X_{r})\,P_{phys}+\text{divergences} \\
&=P_{phys}\, \tilde{T}_{n}(X)\, P_{phys}+\text{divergences}\, ,
\label{eq:3.8}
\end{split}
\end{equation}
because of the formula for the inductive construction of $\tilde{T}_{n}(X)$~\cite{scha}. By means of pseudo-unitarity,
$T_{n}^{\mathrm{K}}(X)=\tilde{T}_{n}(X)$, we arrive at
\begin{equation}
\tilde{T}_{n}^{P}(X)=P_{phys}\, T_{n}^{\mathrm{K}}(X)\, P_{phys} +\text{divergences}\, .
\label{eq:3.9}
\end{equation}
Since on $\mathcal{F}_{phys}$ the K-conjugation agrees with  the $\dagger$-adjoint, we obtain
\begin{equation}
\tilde{T}_{n}^{P}(X)=P_{phys}\, T_{n}^{\dagger}(X)\, P_{phys} +\text{divergences}\, ,
\label{eq:3.10}
\end{equation}
which is the desired form of perturbative unitarity on the physical subspace. Roughly speaking, we have shown
that $\big( P_{phys}\, S(g)\, P_{phys}\big)^{-1}$, constructed by means of the $\tilde{T}_{n}^{P}$'s,
agrees with $\big( P_{phys}\, S(g)^{\dagger}\, P_{phys}\big)$, obtained by means of 
the $T_{n}^{\dagger}$'s, up to divergence terms that do not contribute in the adiabatic limit.

\subsection{Fock Space Decomposition and Gauge Charge}\label{sec:lemma}

The most difficult step is the elimination of the internal projections  on $\mathcal{F}_{phys}$ 
in Eq.~(\ref{eq:3.7}). In order to accomplish this step, we take advantage of the relation
between gauge charge $Q$ and decomposition of the Fock space $\mathcal{F}$,~\cite{asd}.
Since $Q$ and $Q^{\dagger}$ are unbounded operators (the unboundedness is due not only to the presence
of emission and absorption operators but also to  the presence of $\omega(\bol{p})=|\bol{p}|$ in the
expressions for $Q$ and $Q^{\dagger}$ in momentum space, Eq.~(\ref{eq:73})), we have the direct 
decomposition of the Fock space $\mathcal{F}$  in the form:
\begin{equation}
\mathcal{F}=\overline{\mathrm{ran}\; Q}\oplus \ker Q^{\dagger}=\overline{\mathrm{ran}\; Q^{\dagger}}\oplus
\ker Q \, ,
\label{eq:3.11}
\end{equation}
where the over-lining denote closure. Since $Q^2=0$, $\mathrm{ran}\; Q\perp\mathrm{ran}\; Q^{\dagger}$, 
therefore
\begin{equation}
\mathcal{F}=\overline{\mathrm{ran}\; Q}\oplus\overline{\mathrm{ran}\; Q^{\dagger}}\oplus 
\Big( ( \ker Q )\cap (\ker Q^{\dagger}) \Big) \, .
\label{eq:3.12}
\end{equation}
The range of $Q$ and $Q^{\dagger}$ certainly consists of unphysical states only, because $Q$ and $Q^{\dagger}$
only contain emission operators for these unphysical states (absorption operators also appear, but since
these latter  are always paired with the corresponding creation operators, the resulting state is always unphysical).
Therefore, since
\begin{equation}
\Big( ( \ker Q )\cap (\ker Q^{\dagger}) \Big)=\ker \Big( \big\{ Q, Q^{\dagger} \big\} \Big)
=\mathcal{F}_{phys}\, ,
\end{equation}
we find the direct orthogonal decomposition
\begin{equation}
{\mathbf 1}=P_{Q}+P_{Q^{\dagger}}+P_{phys}\, ,
\label{eq:3.13}
\end{equation}
where $P_{Q}$ and $P_{Q^{\dagger}}$ are the projectors onto $\mathrm{ran}\; Q$ and onto
$\mathrm{ran}\; Q^{\dagger}$, respectively. With $K:=\{Q,Q^{\dagger}\}$ positive self-adjoint on 
$\mathcal{F}_{phys}^{\perp}$, there exists $K^{-1}$ on $\mathcal{F}_{phys}^{\perp}$ with $K^{-1}P_{phys}=0$
and $K\,K^{-1}=P_{Q}+P_{Q^{\dagger}}$ so that
\begin{equation}
{\mathbf{1}}=K\,K^{-1}+P_{phys}=P_{phys}+Q\,Q^{\dagger}\,K^{-1} + Q^{\dagger}\, Q \, K^{-1}\, .
\label{eq:3.14}
\end{equation}
If $X_1$ and $X_2$ are two disjoint sets of points, then
\begin{equation}
\begin{split}
P_{phys}\,T_{n_{1}}(x_{1})\,T_{n_{2}}(x_{2})\,P_{phys}&= P_{phys}\,T_{n_{1}}(x_{1})\,\Big[
P_{phys} + Q\, Q^{\dagger}\, K^{-1}+\\
&\quad\qquad +  Q^{\dagger}\,Q\,K^{-1}\Big]\,T_{n_{2}}(x_{2})\,P_{phys}=\\ 
&=P_{phys}\,T_{n_{1}}(x_{1})\,P_{phys}\, T_{n_{2}}(x_{2})\,P_{phys}+\\
&\quad\qquad +P_{phys}\,T_{n_{1}}(x_{1})\,Q\, Q^{\dagger}\, K^{-1}\, T_{n_{2}}(x_{2})\,P_{phys}+\\
&\quad\qquad +P_{phys}\,T_{n_{1}}(x_{1})\,Q^{\dagger}\, Q\, K^{-1}\, T_{n_{2}}(x_{2})\,P_{phys}\, .
\label{eq:3.15}
\raisetag{25mm}
\end{split}
\end{equation}
Since $P_{phys}\,Q=0$, the second term becomes
\begin{gather}
P_{phys}\,T_{n_{1}}(x_{1})\,Q\, Q^{\dagger}\, K^{-1}\, T_{n_{2}}(x_{2})\,P_{phys}=\nonumber\\
=P_{phys}\,\underbrace{\big[T_{n_{1}}(x_{1}),Q\big]}_{\text{divergence}}\,Q^{\dagger}\, K^{-1}\, 
T_{n_{2}}(x_{2})\,P_{phys}=\text{divergence}\, .
\label{eq:3.16}
\end{gather}
Analogously, since $[K^{-1},Q]=0$, the third term becomes
\begin{gather} 
P_{phys}\,T_{n_{1}}(x_{1})\,Q^{\dagger}\, Q\, K^{-1}\, T_{n_{2}}(x_{2})\,P_{phys}=\nonumber\\
=P_{phys}\,T_{n_{1}}(x_{1})\,Q^{\dagger}\,K^{-1}\, Q\,  T_{n_{2}}(x_{2})\,P_{phys}=\nonumber\\
=P_{phys}\,T_{n_{1}}(x_{1})\,Q^{\dagger}\,K^{-1}\,
\underbrace{\big[Q,T_{n_{2}}(x_{2})\big]}_{\text{divergence}}\, P_{phys}=\text{divergence}\, .
\label{eq:3.17}
\end{gather}
Therefore we obtain the relation
\begin{equation}
P_{phys}\,T_{n_{1}}(x_{1})\,T_{n_{2}}(x_{2})\,P_{phys}=
P_{phys}\,T_{n_{1}}(x_{1})\,P_{phys}\, T_{n_{2}}(x_{2})\,P_{phys}+\text{divergences}\, .
\label{eq:3.18}
\end{equation}
The above relation  can be generalized to the case of $r$ $n_{r}$-point  distributions $T_{n_{r}}$ and
used to eliminate the internal physical projectors from Eq.~(\ref{eq:3.7}).

\section*{Acknowledgements}

I would like to thank Prof.~G.~Scharf for his continuous and patient support, Adrian M\"uller and Mark
Wellmann  for stimulating discussions and all the people at the Institute for Theoretical Physics
of the Z\"urich University for the friendly atmosphere.

\end{document}